\documentclass[twocolumn]{svjour3}             
\smartqed  
\usepackage{graphicx,url}
\usepackage[square,authoryear]{natbib}
\usepackage{comment}
\usepackage[english]{babel}
\usepackage{amsmath}
\usepackage{xcolor}
\usepackage{ulem}
\usepackage{siunitx}
\usepackage{soul}

\usepackage{enumitem}
\setlist{parsep=0pt,listparindent=\parindent}
\begin{document}

\title{Backscattered solar Lyman-$\alpha$ emission as a tool for the heliospheric boundary exploration}

\titlerunning{Lyman-$\alpha$ emission as a tool for heliospheric exploration}

\author{
Igor Baliukin 
\and Jean-Loup Bertaux
\and Maciej Bzowski
\and Vladislav Izmodenov
\and Rosine Lallement
\and Elena Provornikova
\and Eric Qu\'emerais
}


\institute{
I. I. Baliukin, V. V. Izmodenov \at
Space Research Institute of Russian Academy of Sciences, Profsoyuznaya Str. 84/32, Moscow 117997, Russia\\
Lomonosov Moscow State University, Moscow Center for Fundamental and Applied Mathematics, GSP-1, Leninskie Gory, Moscow 119991, Russia\\
\email{igor.baliukin@gmail.com}
\and
J.-L. Bertaux, E. Qu\'emerais, \at
LATMOS/IPSL, UVSQ, France
\and
M. Bzowski \at 
Space Research Centre, Polish Academy of Sciences, Warsaw, Poland\\
\email{bzowski@cbk.waw.pl}
\and
R. Lallement \at
GEPI, Observatoire de Paris, PSL University, CNRS, 5 Place Jules Janssen, 92190, Meudon, France
\and
E. Provornikova \at
Johns Hopkins University Applied Physics Laboratory, USA
}

\date{Received: date / Accepted: date}

\maketitle

\begin{abstract}

This review summarizes our current understanding of the outer heliosphere and local interstellar medium (LISM) inferred from observations and modeling of interplanetary Lyman-$\alpha$ emission. The emission is produced by solar Lyman-$\alpha$ photons (121.567 nm) backscattered by interstellar H atoms inflowing to the heliosphere from the LISM. Studies of Lyman-$\alpha$ radiation determined the parameters of interstellar hydrogen within a few astronomical units from the Sun. The interstellar hydrogen atoms appeared to be decelerated, heated, and shifted compared to the helium atoms. The detected deceleration and heating proved the existence of secondary hydrogen atoms created near the heliopause. This finding supports the discovery of a Hydrogen Wall beyond the heliosphere consisting of heated hydrogen observed in HST/GHRS Lyman-$\alpha$ absorption spectra toward nearby stars. The shift of the interstellar hydrogen bulk velocity was the first observational evidence of the global heliosphere asymmetry confirmed later by Voyager in situ measurements. SOHO/SWAN all-sky maps of the backscattered Lyman-$\alpha$ intensity identified variations of the solar wind mass flux with heliolatitude and time. In particular, two maxima at mid-latitudes were discovered during solar activity maximum, which Ulysses missed due to its specific trajectory. Finally, Voyager/UVS and New Horizons/Alice UV spectrographs discovered extraheliospheric Lyman-$\alpha$ emission. We review these scientific breakthroughs, outline open science questions, and discuss potential future heliospheric Lyman-$\alpha$ experiments.

\end{abstract}


\section{Introduction}

The first measurements of ultraviolet radiation in the Lyman-$\alpha$ line (with a wavelength at the center $\lambda_0$ = 121.567 nm) were made at the very beginning of the space era using night rocket launches. In particular, \citet{kupperian1959} launched at night a rocket reaching 146 km of altitude, which allowed to look both to the nadir and the upper hemisphere. The experiment recorded a diffuse Lyman-$\alpha$ emission from all directions. The emission from the nadir hemisphere was interpreted correctly as hydrogen (H) atoms in the atmosphere below the rocket, being illuminated by the (stronger) emission coming from the upper hemisphere. This sky emission displayed a minimum right in the anti-solar direction, $\sim$50$^\circ$ away from the zenith, which could not be explained by the expected radiation from nearby stars. It was (incorrectly) assigned to H atoms in the interplanetary medium scattering solar photons, with the density estimate 0.2 cm$^{-3}$ \citep{kupperian1959}. \citet{shklovsky1959} and \citet{brandt1959} considered two hypotheses concerning a source of the observed emission: (1) H atoms in interplanetary space and (2) H atoms in the Earth's geocorona. The ``geocorona'' term for extended terrestrial atmosphere of H atoms was proposed by \citet{shklovsky1959}. Both papers ruled out the second explanation and concluded that the emission is a result of scattering of solar photons from H atoms in the interplanetary space. Conversely, the following year, \citet{johnson1960} favored the second explanation. Their model suggested an extended exosphere with H atom densities reaching 10 cm$^{-3}$ at the distance $7$ Earth Radii ($R_{\rm E}$).

Further progress was made by using a hydrogen absorption cell (H cell) in Lyman-$\alpha$ radiation measurements during nighttime rocket launches \citep{morton1962}. It was found that 85\% of the measured radiation was absorbed by the H cell. This implied a small Doppler shift relative to the center of the line, and, therefore, proved its terrestrial origin (explanation 2).
The remaining non-absorbed 15\% of the radiation had a large Doppler shift, meaning that such scattered photons were produced by the interaction of a solar photon with a H atom having a high speed relative to the Earth. An atom can have a high speed in two cases: (1) if the flow of atoms moves fast relative to the Earth, and (2) if the scattering gas has a high temperature.
\citet{morton1962} argued that if the radiation was coming from interplanetary H atoms, its absorption by the H cell  would be changing depending on a look direction because of a variable Doppler shift with the angle to the Earth’s orbital motion. Since the 15\% residual intensity did not change with a look direction on the sky, they favored the other explanation, namely that the emission was produced by scattering of solar photons on ``hot'' H atoms (7000 K) in the Earth's upper atmosphere (exosphere). In retrospect, we understand what happened. The H cell was launched on April 17. At this time of year, the Earth’s orbital velocity vector is oriented opposite to the interstellar H flow velocity vector. Hence, the emission from the interplanetary H atoms has a large Doppler shift. A region in the sky where the H cell could have absorbed the interplanetary emission (along the Zero Doppler Shift Circle, see Section \ref{swan_hcell_data}) is quite narrow and was out of lines of sight of the H cell in these rocket experiments.

After having discarded an existence of the extended geocorona in 1959, \citet{chamberlain1963} returned to this concept motivated by the observation of 85\% absorption with the H cell. Extending the work of \citet{johnson1960}, he established a theory of H atom distribution in the exosphere. Now well-known, the Chamberlain's model introduced a concept of escaping, ballistic and satellites particles (H atoms), populating an extended exosphere (geocorona).

Meanwhile, \citet{patterson1963} interpreted the residual 15\% of emission as being due to H atoms in the interplanetary space.  They put forward two hypotheses that the H atoms were either a ``cold'' component of interstellar atoms approaching the Sun or produced at the solar wind (SW) termination shock by charge exchange with neutral galactic (interstellar) H atoms. 
They rejected (wrongly) the ``cold'' component explanation for the same reason as \citet{morton1962}, i.e. a lack of variation in the absorption with changing line of sight (LOS). They favored the latter idea, extending a work of \citet{axford1963} about the termination of the solar wind. A population of atoms produced at the TS is similar to so-called ``hot'' population, which is discussed below in Section 4. From the non-absorbed Lyman-$\alpha$ intensity (4.1 $\times$ 10$^{-4}$ erg cm$^{-2}$ s$^{-1}$ sr$^{-1}$), they derived a distance to the termination shock of 20 AU and a magnitude of the interstellar magnetic field of 3 nanoteslas (nT). Note, these early quantitative estimates resulted from the understanding that the extraterrestrial Lyman-$\alpha$ emission carry imprints of the interaction of the SW with the interstellar medium (ISM).

New experimental data obtained in the 1960s on the Soviet interplanetary spacecraft Zond 1, Venera 2, 3 and 4 \citep{kurt_gemagenova1967, kurt_sunyaev1967} have confirmed the conclusion that 15 \% of the radiation measured during night rocket launches has an extraterrestrial origin. However, it was also found that the maximum of backscattered radiation is located in the direction close to the center of our galaxy \citep{kurt_sunyaev1967}. This prompted new theories about the Galactic sources of observed Lyman-$\alpha$ radiation, which later turned out to be erroneous.

Another breakthrough came in the early 1970s with new measurements of backscattered Lyman-$\alpha$ radiation on the OGO-5 (Orbiting Geophysical Observatory number 5) spacecraft \citep{bertaux1971, thomas1971}. These measurements made it possible to resolve the question on the nature of the source of radiation. The apogee of the OGO-5 spacecraft was outside the geocorona, and the spacecraft itself was purposely put in a rotation mode, which allowed constructing complete maps of the emission intensity. The so-called parallax effect (angular shift of the position of the radiation maximum at different points of the spacecraft's orbit) was discovered, which would be negligible in the case of a Galactic source. Thus, it was established that the source of the measured Lyman-$\alpha$ radiation is indeed interstellar H atoms that flow to the solar system from the ISM.

From the late 1970s until the end of the 1980s, the main results on backscattered Lyman-$\alpha$ radiation were obtained using measurements on Soviet spacecraft Mars-7 \citep{bertaux1976} and Prognoz-5/6 \citep{bertaux1977, bertaux1985, lallement1984, lallement1985b}. These spacecraft were equipped with photometers with an H absorption cell, which made it possible to study not only the absolute value of the radiation (intensity) but also the spectral properties of the line. Backscattered Lyman-$\alpha$ radiation has been measured in the outer heliosphere by Mariner 10 \citep{ajello:78, kumar_broadfoot_1978, kumar_broadfoot_1979}, Pioneer 10 \citep{wu1988, gangopadhyay1989}, Pioneer Venus \citep{ajello_etal:87, ajello_etal:90}, Galileo/Ultraviolet Spectrometer (UVS) \citep{ajello_etal:93}, and Voyager/UVS. Since 1990, it has been measured on Solar and Heliospheric Observatory (SOHO)/Solar Wind ANisotropy (SWAN), Hubble Space Telescope (HST)/Goddard High Resolution Spectrograph (GHRS), Space Telescope Imaging Spectrograph (STIS), and New Horizons/Alice. As we discuss in this review, in the last 30 years, backscattered Lyman-$\alpha$ radiation was observed on several spacecraft and actively studied. A large amount of experimental data has been accumulated, providing us new important knowledge on the heliosphere.

This paper is structured as follows. 
Section 2 describes basic theoretical ideas. 
Section 3 summarizes photometric (intensities) observations. 
Section 4 presents spectrometric and H cell (spectral line shape) observations.
Conclusions, open questions and future experiments are discussed in Section 5.

\section{Basic theoretical ideas}

\subsection{Lyman-$\alpha$ emission}

The Sun constantly emits photons at different frequencies, while the spectrum of solar radiation consists of various selected lines. The solar spectrum in the Lyman-$\alpha$ line has the shape $s(\lambda)$ with a dip in the middle since photons from the center of the line are actively scattered in the atmosphere of the Sun \citep[see, e.g.,][]{lemaire1998, lemaire2005}, so the flux of photons at the Earth's orbit with the wavelength $\lambda$ can be represented as follows:
\begin{equation}
F_{\rm E}(\lambda) = s(\lambda) \cdot F_{\rm E,0}.
\end{equation}
These quantities vary with solar activity and the statistical relationship between total flux in the Lyman-$\alpha$ line $F_{\rm E,0}$ and flux at line center $f$ established by \citet{emerich2005} from Solar Ultraviolet Measurements of Emitted Radiation (SUMER) on board Solar and Heliospheric Observatory (SOHO) can be used to derive the flux at line center:
\begin{equation}
	\frac{f}{10^{12}\: \rm s^{-1}\: cm^{-2}\: nm^{-1}} = 0.64 \left(\frac{F_{\rm E,0}}{10^{11}\: \rm s^{-1}\: cm^{-2}} \right)^{1.21} \pm 0.08
\end{equation}
The daily total solar Lyman-$\alpha$ flux measured at the Earth's orbit ($F_{\rm E,0} \sim 3-7 \times 10^{11}$ photons $\rm s^{-1} cm^{-2}$) is available in the form of so-called solar composite Lyman-$\alpha$ flux, with the most recent version given by \citet[][see \url{http://lasp.colorado.edu/lisird/data/composite_lyman_alpha/}]{machol2019}, which is \\ based on observations from several spacecraft. The variation of the solar Lyman-$\alpha$ line profile with the level of solar activity, represented by the total solar Lyman-$\alpha$ flux, was studied by \citet{tarnopolski_bzowski:09a, kowalska2018a, kowalska2018b, kowalska2020, kretzschmar2018}.

On the way through the heliosphere, the Lyman-$\alpha$ photon can experience scattering. Considering $n$ photons with the same direction $\mathbf \Omega$, wavelength $\lambda$ within ${\rm d} \lambda$ and going through a medium with local hydrogen density $n_{\rm H}$, the absorption probability after crossing a distance ${\rm d} s$ is proportional to the density multiplied by ${\rm d}s$. The absorbed fraction ${\rm d}\tau = -{\rm d} n / n$ is the following:
\begin{equation}
{\rm d}\tau = \sigma (\lambda) n_{\rm H} {\rm d}s,
\end{equation}
where $\sigma({\lambda})$ is the absorption cross-section at wavelength $\lambda$, which has the dimension of an area. The cross-section $\sigma(\lambda)$ is proportional to the projection of the velocity distribution function of H atoms on the direction $\mathbf \Omega$ \citep{quemerais2006a}. To be more precise,
\begin{equation}
\sigma(\lambda) = k_{0}\cdot \hat{f}_{\rm p}(u), \label{eq:cross_section}
\end{equation}
where $\hat{f}_{\rm p}(u)$ -- the normalized projection of H velocity distribution function such as $\int_{-\infty}^{+\infty} \hat{f}_{\rm p}(u)\, {\rm d} u = 1$, and $u = c (\lambda/\lambda_{0} - 1)$ is the value of projected velocity. The total (integrated) cross-section $\sigma_{\lambda}$ can be obtained from the harmonic oscillator theory, and it is given by \citet{mihalas1978}:
\begin{equation}
\sigma_{\lambda} = \int_{0}^{+\infty}\sigma({\lambda})\,{\rm d}\lambda = \left(\frac{\lambda_0^2}{c}\right) \frac{\pi\,e^{2}}{m_{e}\,c}f_{12}  \\
 = 5.445 \times 10^{-15} \rm cm^2 \: \mbox{\normalfont\AA},
\end{equation}
where $m_{e}$ -- mass of electron, and $f_{12} = 0.416$ -- oscillator strength. The proportionality coefficient in Equation \ref{eq:cross_section} is $k_{0} = (c / \lambda_{0}) \cdot \sigma_{\lambda}$, since  ${\rm d} u = (c / \lambda_0) \cdot {\rm d}\lambda$. 

In the interplanetary medium, all absorptions of Lyman-$\alpha$ photons by H atoms are followed by the emission of new photons (resonance scattering). During the process of absorption, the photon energy is transferred to the electron, the electron transits to the second electron orbit (i.e., the atom is promoted to the excited state) and then, since this state is not stable, the electron returns to the ground state, as a result of which a new Lyman-$\alpha$ photon is emitted. In general, the direction of motion and frequency of the original and emitted photons are different. In principle, the direction of movement of the emitted photon can be arbitrary, however, some directions are more probable. A function that determines a probability that the scattering occurs exactly at the angle $\omega$ between the directions of motion of the incident and scattered photons is called the scattering phase function, and it expresses the relation between the absorbed and the emitted intensities. \citet{brandt1959} obtained an expression for the normalized scattering phase function in the Lyman-$\alpha$ line:
\begin{equation}\label{eq:phase_function}
\varphi(\omega) = \frac{11/12+1/4\,\cos^{2}\omega}{4\pi}.
\end{equation}
More recently, \citet{brasken1998} computed a new function based on the general theory of resonance scattering with quantum mechanics. Their result is numerically very similar to the expression \ref{eq:phase_function}.

In the process of scattering, an atom can lose part of its energy, while the frequency of the emitted photon in the coordinate system of the atom will differ from the frequency of the primary photon. This effect is called the natural broadening of the line. However, in the case of heliospheric studies, the natural broadening can be neglected \citep{quemerais2006a}. It can be explained by the fact that the lifetime of an electron at the upper energy level (i.e., when the atom is in an excited state) is small in comparison with the characteristic times of other considered processes, so the change in the velocity of the atom during this time can be neglected. 
Thus, as a result of the interaction of a solar photon with a H atom, a new scattered Lyman-$\alpha$ photon is formed, and the characteristics of the Lyman-$\alpha$ emission are strongly dependent on the velocity distribution function of H atoms. Such scattered emission can be measured by various spacecraft.

The interaction of Lyman-$\alpha$ photon with a H atom is twofold. On one hand, the wavelength of the backscattered Lyman-$\alpha$ photon depends on the velocity of the H atom. On the other hand, when an atom absorbs a photon, the value of atom momentum increases (by photon momentum transfer) in the antisolar direction. Due to subsequent photon emission, it also changes by almost the same amount, but in a random direction. The average momentum change of many emissions is zero, since the scattering phase function is symmetrical along the photon motion direction (see Equation \ref{eq:phase_function}). Therefore, a net effect of many scatterings can be described as a radiation pressure force that acts on the atom in the antisolar direction. 

The radiation pressure force is directly proportional to the solar Lyman-$\alpha$ flux that in the optically thin medium approximation (no absorption by atoms) falls off $\propto1/r^2$ with distance to the Sun, as the solar gravitational force. The ratio $\mu$ of solar radiation pressure and solar gravitational force is commonly used in the literature. The dependence of the radiation pressure on time, radial velocity, and heliolatitude of the H atom was studied by \citet{tarnopolski_bzowski:09a, kowalska2018a, kowalska2018b, kowalska2020}.
It is important to note that, strictly speaking, the optically thin medium approximation is not applicable in the whole heliosphere since the hydrogen number density increases with distance from the Sun. The interplanetary medium can be considered optically thin only until 10 AU \citep{quemerais2000}. Therefore, the solar radiation pressure and the ratio $\mu$ depend also on the distance to the Sun. Due to the absorption of photons by H atoms, the solar pressure drops more rapidly than $\propto1/r^2$. While the importance of absorption increases with distance from the Sun, the hydrogen distribution is essentially affected by the solar pressure (and gravitation) only in the vicinity of the Sun. A step towards the exploration of the influence of absorption of the solar Lyman-$\alpha$ photons on the hydrogen distribution is the recent analysis by \citet{kowalska2022} who showed that ignoring the absorption in total results in underestimating the hydrogen number density (in the worst case by $\sim$9\% in the downwind direction). 

Actually, the secondary (scattered) photons can also provide the momentum transfer to the H atoms, so the regions of enhanced number density ``push'' other atoms away, even though the influence should be minor. Since the hydrogen distribution depends on Lyman-$\alpha$ emission and vice versa, to investigate the correct influence, one should perform self-consistent numerical calculations of the atom interaction with Lyman-$\alpha$ emission (with the multiple scattering effect taken into account).

The main source of the heliospheric Lyman-$\alpha$ emission is the backscattered solar photons by the interstellar H atoms, which penetrate the heliosphere due to the relative motion of the Sun and LISM (Local InterStellar Medium). Actually, astronomers are using the acronym ``LISM'' to designate the interstellar medium, which is not far from the Sun. Later on, the acronym ``VLISM'' (Very Local InterStellar Medium) was used to designate the particular part of the LISM in which is presently located the Sun, and which interacts with the heliosphere. For simplicity, in the following we use ``LISM'' instead of ``VLISM''. Numerous measurements of the backscattered solar Lyman-$\alpha$ radiation have motivated the development of theoretical models of interstellar H atoms propagation from the LISM to the vicinity of the Sun. The next Section \ref{sec:Hdistribution} briefly reviews existing models of hydrogen distribution in the heliosphere.

To model the interplanetary (IP) Lyman-$\alpha$ background, one should (a) calculate the hydrogen distribution in the heliosphere, and (b) solve the radiative transfer equation. To obtain the correct solution of the radiative transfer equation, the multiple scattering effects must be considered everywhere in the heliosphere, which has been proven quantitatively by \citet{quemerais2000}. However, since the problem is computationally time-consuming, some approximations, which take into account only single scattered photons, have been developed:
(1) the optically thin (OT) approximation neglects the extinction of photons due to the absorption, and it generally provides appropriate intensity estimates in the inner heliosphere since the extinction and multiple scattering effects act to balance each other;
(2) the primary term (PT) approximation gives the exact single scattering intensity when both extinctions (between Sun and the scattering point, and on the LOS between the scattering point and the observer) are included; in the inner heliosphere, the PT approximation gives $\sim$60\% of the intensity seen from 1 AU \citep{quemerais2000};
(3) the self-absorption (SA) approximation proposed by \citet{bertaux1985} is the trade-off between the OT and PT cases; in contrast to the OT approximation, the extinction on the LOS between the scattering point and the observer is included in this case. Numerical estimates of the errors introduced by these approximations are provided by \citet{quemerais2000}. Details on the simulations of the interplanetary hydrogen Lyman-$\alpha$ background can be found in \citet{quemerais2000, quemerais_izmod_2002, quemerais2008}.

\subsection{Hydrogen distribution in the heliosphere}\label{sec:Hdistribution}

\citet{fahr1968} and \citet{blum1970} proposed a so-called ``cold model'', which assumes that all H atoms have the same velocity in the LISM. However, measurements of the interplanetary Lyman-$\alpha$ background with the use of the hydrogen absorption cell on board the Prognoz-5 spacecraft allowed to estimate the LISM temperature, which appeared to be not negligible \citep[$\sim$8000 K; see, e.g.,][]{bertaux1977}, so the thermal and bulk flow speeds of atoms in the LISM are comparable. Further developments were associated with consideration of the realistic temperature and the corresponding thermal velocities of H atoms in the LISM, which resulted in construction of so-called ``hot models''. In the classical formulation of a ``hot model'', the problem is stationary and axisymmetric, and the velocity distribution function of H atoms at infinity (i.e., in the LISM) is a uniform Maxwell-Boltzmann distribution. \citet{meier1977} and \citet{wu1979} presented the analytical solution for the interstellar hydrogen velocity distribution. The solution takes into account the solar gravitational force, the solar radiation pressure, and the losses due to photoionization and charge exchange with the solar wind protons.

Further improvements can be in practice divided into two groups:
\begin{enumerate}
    \item Development of models, which take into account temporal and heliolatitudinal variations of the solar wind. Temporal variations are caused by the 11-year cycle of solar activity and have been considered in many works \citep[e.g.,][]{bzowski1995, summanen1996, bzowski1997, pryor2003, bzowski2008}. The heliolatitudinal variations are connected with the non-isotropic SW structure. \citet{joselyn1975} were the first to show that the anisotropic SW would strongly affect the ISN (InterStellar Neutral) H distribution in the heliosphere. Several authors \citep{lallement1985b, bzowski_etal:02a, bzowski_etal:03a, bzowski:03, pryor2003, nakagawa2009} assumed some analytical relations for the heliolatitudinal variations of the hydrogen ionization rate and took them into account in the frame of the hot model. Additionally, some models \citep{tarnopolski_bzowski:09a, kubiak_etal:21a} take into account the dependence of radiation pressure force on radial velocity of ISN H atoms and the influence of the Lyman-$\alpha$ absorption on the radiation pressure force \citep{kowalska2022}.
    
    \item Development of models, which take into account disturbances of the interstellar hydrogen flow at the heliosheath. Charge exchange of protons \citep{wallis1975} with primary H atoms in the heliospheric interface creates the ``secondary'' population of H atoms, which have the individual velocities of their original parent protons, so the velocity distribution function of newly created atoms depends on the local plasma properties. The accumulation of the heated and decelerated secondary component creates the so-called ``hydrogen wall'' (the region of the pile-up of H atoms between the heliopause and the bow shock) that was firstly reported by \citep{baranov1991}. 
    Such hydrogen walls have been observed around other stars \citep{wood2005}.
    Therefore, a mixture of the primary and secondary interstellar H atoms penetrates the heliosphere. The properties of the H atom component depend on both the LISM parameters and the plasma distribution in the heliosheath. Disturbances of the interstellar hydrogen flow in the heliosheath were included in the hot model using the different approaches suggested by \citep{bzowski_etal:08a, nakagawa2008, katushkina2010, izmod2013}. 

\end{enumerate}

Modern versions of the hot model of the interstellar hydrogen distribution in the heliosphere, such as the models by \citet{katushkina2015a} and \citet{kubiak_etal:21a, kubiak_etal:21b}, take into account both the solar cycle and heliolatitudinal variations of the solar wind and the disturbances of the hydrogen flow by the heliospheric interface.

\citet{kleinmann_sameissue} presents an overview of the current state of research in global modelling of the heliospheric interface. \citet{galli_sameissue} summarizes all direct observations of interstellar H atoms (and other neutrals), most of which were acquired with the Interstellar Boundary Explorer (IBEX) launched in 2008.

\section{What do we learn from maps and radial/angular profiles of backscattered Lyman-$\alpha$ emission?}

\subsection{SOHO/SWAN sky maps from 1 AU}

From 1996 to the present, the backscattered solar Lyman-$\alpha$ emission has been measured by the SWAN instrument (Solar Wind ANisotropy) on board the SOHO spacecraft placed in a halo orbit around Lagrange point $\rm L_1$. The SWAN instrument was designed and built by the Service d'Aéronomie (France) and the Finnish Meteorological Institute (Finland). Consisting from the two identical Sensor Units, the instrument is equipped with a two-axis periscope, which allows obtaining a complete map of the sky in Lyman-$\alpha$ intensity in about 24 hours \citep{bertaux1995}. Each Sensor Unit (SU) covers roughly one half of the sky: SU+Z views the ecliptic Northern Hemisphere and SU+Z -- the Southern Hemisphere,  with some overlap between the two hemispheres. A recent comparison of the SWAN data (from 1996 to 2018) with model calculations shows a sensitivity degradation of the SWAN Sensor Units with the rate of $\sim$2\% per year. In addition, SWAN is equipped with a hydrogen cell, which makes it possible to study the Lyman-$\alpha$ line shape and derive spectral properties. The results obtained with the usage of the SWAN H cell are discussed in Section \ref{swan_hcell_data}.

\begin{figure*}
\centering
\includegraphics[width=1.0\textwidth]{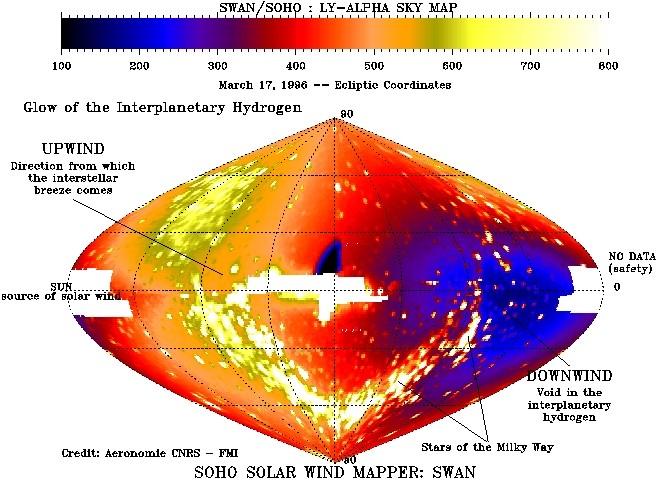}
\caption{The full sky map (in ecliptic J2000 coordinates) of the Lyman-$\alpha$ intensity observed by SWAN/SOHO on 17 March 1996. The white areas represent regions of the sky which can not be observed by SWAN or which are contaminated by hot stars of the Galactic plane. The color represents the intensity in counts per second per pixel (one square degree). Credit: \protect\url{https://sohowww.nascom.nasa.gov/gallery/Particle/swa005.html}.
}
\label{fig:swan_fullsky}
\end{figure*}

Figure \ref{fig:swan_fullsky} shows a typical full-sky map of Lyman-$\alpha$ intensity (in the solar ecliptic coordinates) obtained by SWAN/SOHO on 17 March 1996. The color represents the intensity in counts per second per pixel (one square degree), which corresponds to 1.3 Rayleigh (1 Rayleigh = 1 R = 10$^6$/4$\pi$ photons cm$^{-2}$ s$^{-1}$ sr$^{-1}$). A maximum of Lyman-$\alpha$ intensity is seen in the upwind direction, while in the opposite direction the emission is much weaker since in the downwind region the number density of H atoms is smaller (so-called ionization cavity). The white areas represent regions of the sky that are  masked for safety reasons: the area around the Sun (with $\approx 0^\circ$ ecliptic longitude), the anti-solar region (near the center of the map), which is contaminated by parts of SOHO spacecraft body (illuminated by the Sun). There is also a shark fin-shaped piece of hardware above the anti-solar direction, which is placed to protect the SU+Z from SOHO thrusters firings. These three artificial features associated with the spacecraft (Sun shade, anti-solar contaminated region, shark fin) are moving from right to left with $\approx 1^\circ$ shift per day since the SOHO orbit follows the Earth on its yearly orbital motion.

As the name suggests, the main purpose of the SWAN instrument is to study the heliolatitude dependence (or anisotropy) of the solar wind parameters by analyzing the intensity of the backscattered solar Lyman-$\alpha$ radiation. The parameters of the solar wind determine the ionization rate of H atoms due to the charge exchange with the solar wind protons in the heliosphere. In the region of the supersonic solar wind the charge exchange ionization rate $\beta_{\rm ex}(t, {\bf r}) \approx n_{\rm sw} \cdot V_{\rm sw} \cdot \sigma_{\rm ex} (V_{\rm sw})$, where $n_{\rm sw}(t, {\bf r})$ and $V_{\rm sw}(t, {\bf r})$ are the solar wind number density and velocity, and $\sigma_{\rm ex}$ is the charge exchange cross-section \citep[see, e.g.][]{linsday_stebbings2005}. The cross-section generally depends on the relative velocity between the H atom ($\sim$20 km/s) and the solar wind proton ($\sim$400-800 km/s). However, the latter is much higher and for solar wind velocities the dependence of $\sigma_{\rm ex}$ is weak. Therefore, the charge exchange ionization rate is proportional to the solar wind mass flux. If, for example, the charge exchange ionization rate in the plane of the solar equator is greater than at the poles, then the number density of H atoms in the plane of the equator is less than at the poles. This means that the intensity of the solar Lyman-$\alpha$ radiation backscattered by the H atoms will have a minimum in the equatorial plane and increase monotonically towards the poles. Thus, measurements of the backscattered Lyman-$\alpha$ emission can be used as a tool for an indirect diagnostic of the heliolatitude dependence of the solar wind parameters.

\begin{figure*}
\centering
\includegraphics[width=1.0\textwidth]{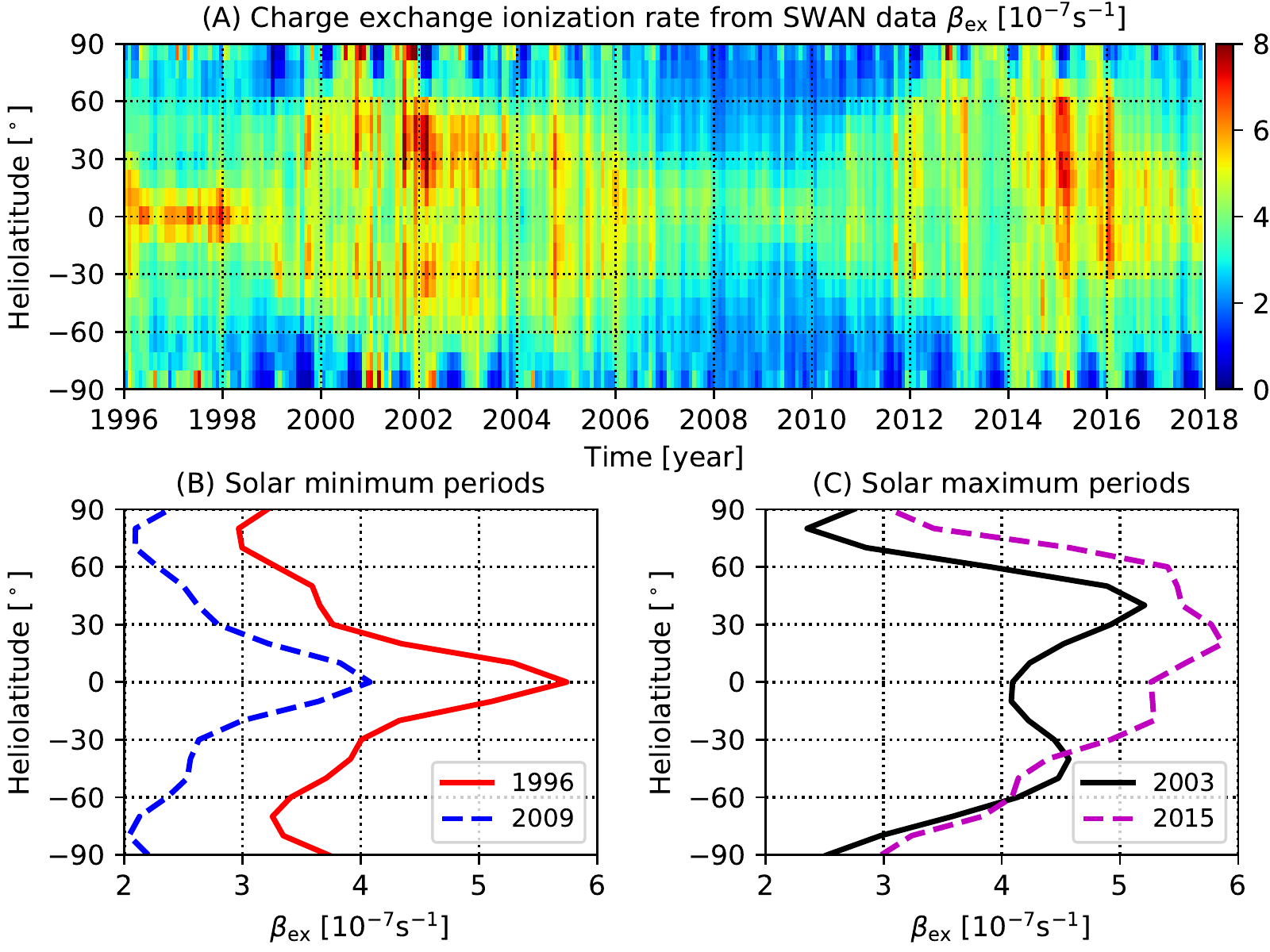}
\caption{(A) Temporal and heliolatitudinal variations of the charge exchange ionization rate derived from SOHO/SWAN data. Panels B and C show averaged latitudinal profiles in particular years of the solar minimum (1996 and 2009) and solar maximum (2003 and 2015) periods, respectively. Adapted from \citet{katushkina2019}.} 
\label{fig:charge_exchange}
\end{figure*}

Indeed, the distribution of Ly-$\alpha$ intensities recorded by Prognoz-5 and 6 along several great circle scans during 1976-77 (solar minimum) displayed a conspicuous dip near the ecliptic plane in the upwind sky region. This was interpreted \citep{bertaux1996} as the trace of a “groove” carved by a peak of enhanced solar wind flux along the heliospheric current sheet, which was quite flat at this time of the solar minimum. This confirmed the idea of \citet{joselyn1975}, except that they had predicted the inverse effect: higher solar flux outside the solar equatorial plane. The effect of the groove is visible on the left side of Figure \ref{fig:swan_fullsky} (upwind region of the maximum emission). The two distinct maxima of the emission are seen in the north and south lobes, which is representative for the solar minimum conditions. A corresponding heliolatitudinal profile of the charge exchange ionization rate is shown in Figure \ref{fig:charge_exchange}B by a red curve.

SWAN has been mapping the sky in the Lyman-$\alpha$ intensity for more than 25 years. Using SWAN data, it is possible to reconstruct a charge exchange ionization rate of H atoms as a function of time and heliolatitude. To do this, it is necessary to simulate a distribution of H atoms and the intensity of backscattered radiation for every moment of time, assuming different dependencies of the charge exchange ionization rate on heliolatitude. Then comparing simulation results with the SWAN data, an inverse problem can be solved to obtain the dependence that fits the data best. An early attempt to reconstruct the solar wind structure using the SOHO/SWAN data was made by \citet{kyrola1998}. This developed technique is described in detail by \citet{quemerais2006b}, and the results are presented in \citet{lallement2010, izmod2013, katushkina2013}.

Using the SOHO/SWAN Lyman-$\alpha$ maps during 1996--2018, \citet{katushkina2019} have studied a temporal and heliolatitudinal dependence of the solar wind mass flux at 1 AU. Figure \ref{fig:charge_exchange} shows the charge exchange ionization rate retrieved from these SWAN observations. During the solar minima (in 1996 and 2009) the charge exchange ionization rate has a distinct maximum at zero latitude (see also Figure \ref{fig:charge_exchange}B). During the solar maxima (2003 and 2015), there are two separated maxima at mid-latitudes ($\pm$30 – 50$^\circ$) in 2003, which seem to be merged in 2015 (see Figure \ref{fig:charge_exchange}C). Unfortunately, these maxima in 2003 were not observed by Ulysses (operated in 1990 – 2010), the only spacecraft that performed in situ measurements out of the ecliptic plane \citep{mccomas2008}, because its trajectory passed between them. In Figures \ref{fig:charge_exchange}A and \ref{fig:charge_exchange}C, it is seen that the latitude distributions are somewhat different for the two successive solar maximum periods. Note, however, that the Sun's magnetic field returns to the same polarity after two 11-year solar cycles (22 years). Continuous sky mapping with the SOHO/SWAN mission will allow determining whether the latitudinal pattern will follow the same trend in the next 22-year cycle.

\citet{katushkina2019} compared latitudinal variations of the solar wind mass flux inferred from the SWAN results and obtained from the  Wang-Sheeley–Arge (WSA)-Enlil model of the inner heliosphere. The model simulates a global 3D solar wind structure from the solar corona to 1 AU driven by the output from the coronal WSA model (WSA-Enlil is available for runs on request at the website of the Community Coordinated Modeling Center, \url{https://ccmc.gsfc.nasa.gov/}). Despite some quantitative differences, the WSA-Enlil results are consistent with the SW latitudinal structure seen in the SWAN data at the solar maxima. However, physical reasons for formation of the two maxima in middle latitudes remain unknown.

\citet{koutroumpa2019} used the same dataset of Lyman-$\alpha$ observations from SOHO/SWAN during 1996-2018 to calculate heliolatitudinal profiles of the SW velocity. To derive the velocity profiles, they assumed that either the dynamic pressure or the energy flux (both calculated using the OMNI data) conserves over heliolatitude. Both methods showed similar results (up to at least $\pm$60$^\circ$). The profiles are consistent with annual averages of the solar wind velocity profiles from Interplanetary scintillation observations provided by \citep{tokumaru2013}, and also with in situ measurements by Ulysses \citep{mccomas2013}. Therefore, the SW latitudinal structure derived from SWAN is in agreement with the assumptions of the dynamic pressure and energy flux invariance. However, during solar maxima, the differences are more pronounced. A possible explanation is that SWAN data indicate global heliospheric conditions, while OMNI and Ulysses data provide in situ observations of the local SW conditions.

\citet{strumik_etal:20a} used a database of International Ultraviolet Explorer (IUE) spectral ultraviolet (UV) observations of astrophysical point sources and SWAN maps to estimate the distribution of non-Lyman-$\alpha$ extraheliospheric emission in the sky. This emission is a result of a contamination by UV continuum of stars, resolved or unresolved. They developed a method to estimate this emission given the spectral characteristics of individual instruments and applied it to SWAN. They found that the extraheliospheric signal is expected at a level of several dozens Rayleigh, peaking at the Galactic disk. Away from the Galactic disk, it is estimated at $\sim 20$ R, but the distribution of the intensity is quite patchy. These results support earlier SWAN sky-maps of this type of stellar contamination, obtained with a dedicated pixel (1$^\circ$ FOV) equipped with a $\rm Ba F_2$ filter, which cuts off photons below 140 nm (thus, excluding Lyman-$\alpha$). Such a map was used as a screening tool when deriving the latitudinal and temporal distribution of the total ionization rate from SWAN sky maps \citep{quemerais2006b}.

\citet{strumik_etal:21b} used SWAN observations performed during 1996-2020 to investigate an anisotropy of the solar Lyman-$\alpha$ emission. They used the WawHelioGlow model of the heliospheric Lyman-$\alpha$ background (helioglow) by \citet{kubiak_etal:21a}, the model of the solar Lyman-$\alpha$ illumination by \citet{kowalska2020}, and a model of charge exchange and photoionization by \citet{sokol_etal:20a}, which in the charge exchange part is based on interplanetary scintillation observations. They assumed that the solar Lyman-$\alpha$ emission is anisotropic in heliolatitude, similar to \citet{pryor_etal:92}. They obtained a good agreement between the model and observations, assuming that the ratio of the solar polar to equatorial emission varies with time from $\sim$0.85 in solar maximum to $\sim$1 during solar minimum. Also, they found that these ratios are not identical for the north and south poles.

In addition to determining the heliolatitudinal and temporal evolution of the solar wind mass flux, SWAN enabled many other important investigations with significant results. One of them is the ability to study evolution of comets by observing Lyman-$\alpha$ radiation backscattered by cometary H atoms \citep{bertaux1997}. From 1996 to 2016, 61 comets were observed, and their $\rm H_2O$ production rate monitored during their passage in the inner solar system \citep[][and references therein]{combi2019}.

\citet{bertaux2000} and \citet{quemerais2002} showed that the sky maps of the Lyman-$\alpha$ intensity obtained by SWAN can be used to detect solar flares, including those which occur on the far side of the Sun. Interplanetary H atoms are like a screen on which solar Lyman-$\alpha$ radiation is scattered. Therefore, if the solar flux from some region on the Sun increases sharply, which usually happens during flares, a noticeable intensity increase of the backscattered radiation will be seen by SWAN in a certain region of the sky. A solar flare, which occurs on the opposite side of the Sun, as seen from the Earth, will face the Earth in about 14 days (half the period of the Sun's rotation around its axis). Thus, SWAN data allow remote diagnostics of the flare location and intensity already up to two weeks before it becomes available for direct viewing at the solar limb from Earth's orbit. These studies are directly relevant to space weather research and of practical interest since strong solar flares are sources of solar energetic particles which can damage spacecraft electronics.

\subsection{Radial and angular profiles of the Lyman-$\alpha$ intensity}

\subsubsection{Voyager/UVS}\label{sec:voyager}

Of great interest are measurements of backscattered emission by Voyager 1 and 2 spacecraft launched in 1977. The Voyagers move radially from the Sun in two different directions in the nose part of the heliosphere ($\sim 35^\circ$ above (V1) and below (V2) the ecliptic plane). 
Voyager 1 \& 2 crossed the termination shock in 2004 and 2007 at distances of $\sim$94 AU \citep{decker2005, stone2005} and $\sim$84 AU \citep{decker2008, stone2008}, respectively. The heliopause crossings from Voyager 1 \& 2 occurred in 2012 and 2018 at $\sim$122 AU \citep{krimigis2013, stone2013, burlaga2013, gurnett2013} and $\sim$119 AU \citep{krimigis2019, stone2019, richardson2019, gurnett2019, burlaga2019}, respectively.

Both spacecraft have the Ultraviolet Spectrometer (UVS) on board, which is a grating spectrograph with photons collected in 128 different channels according to their wavelength in the range 54 -- 170 nm. The Lyman-$\alpha$ signal is obtained by summing the nine spectral channels (from 70 to 78). A detailed description of the instrument can be found in \citet{broadfoot1977}. Note that the emission due to unresolved sources mentioned in the previous section applies to wide-band instruments, and not to Voyager measurements. As a matter of fact, the UVS on board Voyager is a spectrometer, and the Lyman-$\alpha$ emission line is measured above the stellar continuum \citep[see all details in the Supplementary Online Material of][]{lallement2011}. 

While the UVS instrument on Voyager 2 was turned off in 1998, the UVS on Voyager 1 had been working until 2016. However, an abrupt decrease in the signal was noticed in October 2014 (possibly due to issues with the instrument), and since then, the data are more difficult to interpret. The Voyager 1/UVS observations have been performed in three phases:
(1) 1979--1993 ($\sim$7--53 AU), a variety of LOS's is used with some rolls over the sky \citep{lallement1991, hall1993, quemerais2013, fayock2015}; 
(2) 1993--2003 ($\sim$54--89 AU), regular scans over the sky from the upwind to downwind directions \citep{quemerais2009, quemerais2010, lallement2011, katushkina2016}; 
(3) 2003--2014 ($\sim$90--130 AU), all measurements were performed for nearly the same LOS close to the upwind direction \citep{quemerais2009, katushkina2017}.

The dependence of the Voyager UVS upwind intensity with distance to the Sun was studied by many authors based on a power-law description ($I \propto r^\alpha$) to estimate how fast it falls off with heliocentric distance. Assuming a constant density with distance, the power-law coefficient $\alpha$ is expected to vary between $-1$ and $-2$. In the limiting case of the optically thin medium $\alpha = -1$, since the volume emission (emissivity) is proportional to the constant density and the solar flux that is decreasing $\propto 1 / r^2$, and the intensity is the integral of the emissivity over the LOS. However, the optical thickness increases with distance and modifies the power-law coefficient, so the other limit $\alpha = -2$ of the optically thick case (when multiply scattered photons dominate) was estimated by \citet{hall1992}.

The first extensive study of the radial variation of the Lyman-$\alpha$ intensity observed by Voyager UVS in the upwind direction was performed by \citet{hall1993}, based on the hot model of hydrogen distribution. However, the numerical values derived by \citet{hall1993} were not accurate since the data should be corrected for variations of the solar Lyman-$\alpha$ flux, and their correction was based on the old solar flux measurements. Using a new correction for the illuminating Lyman-$\alpha$ flux and a more realistic model of the hydrogen distribution in the heliosphere including effects of the heliospheric interface \citep{baranov1993, izmodenovetal2001}, \citet{quemerais2003} have confirmed the qualitative result obtained by \citet{hall1993}. The analysis performed by \citet{quemerais2003} based on the Voyager 1/UVS data allowed estimating the intensity decrease with power-law coefficient $\alpha = -1.58 \pm 0.02$ for distances between 50 and 65 AU equal, and, more surprising, $\alpha = -0.22 \pm 0.07$ for observations at distances further than 70 AU from the Sun. Therefore, this result indicates that the upwind Lyman-$\alpha$ intensity is decreasing substantially slower than any model predicts (whether it includes the effect of the heliospheric interface or not).

The measurements of the backscattered solar Lyman-$\alpha$ emission on Voyager 1 and 2 in the period since 1993 were analyzed by \citet{quemerais1995, quemerais2003, quemerais2010}. In these works, for the analysis the distributions of H atoms were obtained from the results of the self-consistent model of the heliospheric interface \citep{baranov1993}. The Voyager measurements are of particular interest, since during the period under consideration they were already quite far from the Sun (at a distance of more than 40 AU). At such distances, the influence of local effects associated with the Sun on the distribution of H atoms becomes small, and therefore the intensity of the backscattered Lyman-$\alpha$ emission is mainly determined by the parameters of the atoms at the heliospheric boundary. In particular, \citet{quemerais2010} suggested that a hydrogen wall is ``visible'' in the Voyager data. Namely, in the direction close to the LISM inflow, a region of increased intensity (an emission ``bump'') is observed, presumably associated with the solar Lyman-$\alpha$ photons backscattered by secondary interstellar atoms that form a hydrogen wall at the heliospheric boundary. However, later it turned out that the conclusions of \citet{quemerais2010} are not correct, since the authors found an error in the radiation transfer code \citep{lallement2011}. Theoretical calculations have shown that the maximum intensity present in the measurement data cannot be explained by the effect of the hydrogen wall. 

A detailed re-analysis by \citet{lallement2011} showed that the position of the excess intensity found in the Voyager data, correlates with strong $\rm H_{\alpha}$ emission from so-called HII regions, mostly located along the Galactic plane. This made it possible to conclude the Galactic nature of the excess emission observed by Voyager. The model-dependent estimation of the intensity of the Galactic emission was $\sim$3--4 R (toward nearby star-forming regions). Thus, in 2011 it was first established that the Galactic Lyman-$\alpha$ radiation can be measured from inside the heliosphere. The measurements of the Galactic Lyman-$\alpha$ emission on the Voyager 1 spacecraft open a way for future cartographic and spectral experiments studying the Galactic Lyman-$\alpha$ radiation, and also make it possible to verify the radiation transfer models that are currently used in the study of the properties and the structure of the interstellar medium. Note that this amount of emission is restricted to the ``bump'' of emission seen above the otherwise smoothly varying signal. This means that any uniformly distributed extraheliospheric (Galactic or extragalactic) Lyman-$\alpha$ emission could not be detected in this study and the existence of such an additional signal (see the next paragraph), in addition to the detected signal, is not precluded at all.

\citet{katushkina2016} performed the analysis of the Lyman-$\alpha$ data obtained by Voyager 1 during the spatial scans in 1993--2003 while Voyager 1 was at 53--88 AU from the Sun based on the state-of-the-art self-consistent kinetic-magnetohydrodynamic (kinetic-MHD) model of the heliospheric interface \citep{izmod2015} and radiative transfer model \citep{quemerais2000}. The authors have shown for the first time that the ratio of the Lyman-$\alpha$ intensities in the downwind and upwind directions in the outer heliosphere is sensitive to the configuration (peak value and location) of the hydrogen wall. Since the H-wall is a source of Doppler-shifted backscattered Lyman-$\alpha$ photons, it can be seen from inside the heliosphere, so the Voyager 1/UVS Lyman-$\alpha$ data can be used for remote sensing of the H-wall. The authors show that the heliospheric model by \citet{izmod2015}, which is consistent with many other measurements including Ly-$\alpha$ data from both Voyager 1 and 2 in 1980--1993, provides a systematically larger downwind to upwind intensity ratio compared with the UVS data in 1993--2003, and to decrease the ratio, a higher and/or closer H-wall is needed.

\citet{katushkina2017} presented the Lyman-$\alpha$ intensities measured by Voyager 1/UVS in 2003--2014 (at 90--130 AU from the Sun). The data show an unexpected behavior in 2003–2009: the ratio of observed intensity to the solar Lyman-$\alpha$ flux is almost constant. The authors performed numerical modeling of these data using the heliospheric model \citep{izmod2015}, which predicted (for various interstellar parameters) a monotonic decrease of intensity not seen in the data. Two possible scenarios that explain the data qualitatively were proposed: (1) the formation of a dense ($\sim$10 cm$^{-3}$) layer of H atoms near the heliopause that provides an additional strongly Doppler-shifted (with velocities +50 km/s) backscattered Lyman-$\alpha$ emission, which is not absorbed inside the heliosphere and may be observed by Voyager 1 ($\sim$35 R of intensity from the layer is needed); (2) the existence of an external nonheliospheric Lyman-$\alpha$ component (of Galactic or extragalactic origin), with $\sim$25 R of additional emission (as their parametric study showed). Due to the quite unusual parameters of H atoms in the first scenario, the second scenario appears to be more probable. \citet{katushkina2017} have demonstrated that the additional constant emission of $\sim$15-20 R leads to a good agreement with the Voyager 1 data in 1993-2003 as well, even without a higher/closer H wall suggested by \citet{katushkina2016}. Therefore, the additional constant emission allows explaining the Voyager 1 data not only after 2003, but also for the previous period of observations (1993–2003).

\subsubsection{New Horizons/Alice}\label{sec:new_horizons}

After exploring the Pluto system and the Kuiper Belt object Arrokoth, now the New Horizons mission is on the escape trajectory from the solar system. It provides excellent opportunity for observations of interplanetary Lyman-$\alpha$ emission in the outer heliosphere, and, in particular, to study the Lyman-$\alpha$ background as a function of distance from the Sun. The Alice ultraviolet spectrograph on the New Horizons spacecraft is equipped with a telescope, a Rowland-circle spectrograph, a double-delay-line microchannel plate (MCP) detector at the focal plane, and associated electronics and mechanisms \citep{stern2008}. The wavelength range of the instrument is 52–187 nm with a filled-slit spectral resolution of 0.9 nm.

During the cruise, the Alice ultraviolet spectrograph has performed multiple observations of the interplanetary Lyman-$\alpha$ emission. In late 2007, 2008, and 2010 single-great-circle scan observations were performed during the annual checkouts (ACOs 1, 2, and 4, at 7.6, 11.3, and 17.0 AU from the Sun, respectively). Before the Pluto flyby (at distance $\sim$33 AU in 2015) the observational strategy was changed to provide a denser sky using six great circles spaced 30$^\circ$ apart. The great circles were chosen to avoid directions towards the Sun and, as much as possible, bright UV stars. Two six-great-circle Lyman-$\alpha$ scans of the sky were made just before and after the encounter of Pluto, and the other five scans -- during the Kuiper Belt Extended Mission (KEM) at a cadence of roughly six months in 2017--2020 at distances 38--47 AU from the Sun. The plan for future observations is to continue this periodicity until the crossing of the solar wind termination shock. A detailed description of New Horizons/Alice observations can be found in the recent papers by \citet{gladstone2013, gladstone2021}. 

Besides the comparison with the Voyager 1/UVS data, \citet{katushkina2017} performed a comparison of the model calculations with the New Horizons/Alice data from ACO-2 and ACO-4 rolls (see Section 6 of their paper). To calculate hydrogen distribution in the heliosphere, the global 3D stationary kinetic-MHD model of the SW/LISM interaction by \citep{izmod2015} was employed, and to calculate the solar Lyman-$\alpha$ emission backscattered by H atoms, the radiative transfer code developed by \citet{quemerais2000} was used. The comparison has shown that the model reproduces the data very well everywhere except the upwind region, where the excess brightness is approximately 20 R. Moreover, the value of the additional emission needed to fit the data is very similar for both rolls, which indicates the invariance of this excess with the distance to the Sun. Such an excess over the model supports the findings based on the analysis of the Voyager 1/UVS data, but does not bring the information on the origin of the additional emission.

\citet{gladstone2018} analyzed New Horizons/Alice observations from 2007 to 2017 at distances 7.6–39.5 AU from the Sun and found that falloff of interplanetary Lyman-$\alpha$ emission in the direction close to the upwind can be well approximated by a $1/r$ dependence (where $r$ is the distance from the Sun to the spacecraft) with an additional constant emission of $\sim$40 R. \citet{gladstone2018} concluded that it is a possible signature of the hydrogen wall or Galactic/extragalactic background. This study also showed that New Horizons/Alice data agrees with the earlier Voyager/UVS observations.

\citet{gladstone2021} have analyzed further data, which include six great circles spread over the sky at 30$^\circ$ intervals, from distances up to 47 AU from the Sun. Based on the two independent approaches (fitting of the Lyman-$\alpha$ brightness $1/r$ falloff with distance from the Sun and estimation of the residuals between the observed brightness and model calculations), the distant Lyman-$\alpha$ background of 43 $\pm$ 3 R was evaluated. \citet{gladstone2021} concluded that this excess emission is constant and distributed almost uniformly over the sky, in contrast to reported earlier \citep{gladstone2018} excess only in the upstream direction. Additionally, this analysis showed a weak correlation of the background distribution with the local cloud structure and the absence of signatures of emission, which can be associated with the hydrogen wall.

New Horizons mission continues its journey beyond 50 AU and maps the sky in Lyman-$\alpha$ emission intensity with Alice instrument.

\section{What do we learn from the spectral properties of backscattered Lyman-$\alpha$ emission?}

The interplanetary Lyman-$\alpha$ background seen from 1 AU is mainly affected by the variability of the Sun, since in its vicinity the H atoms are strongly influenced by ionization processes (charge exchange with protons and photoionization), solar radiation pressure and gravitational forces. Therefore, it is challenging to study the effects of the heliospheric interface using photometric observations in the inner heliosphere. Voyager and New Horizons missions obtained unique observations of backscattered Lyman-$\alpha$ emission intensity at large distances from the Sun (see previous Sections \ref{sec:voyager} and \ref{sec:new_horizons}). However, to distinguish contributions into the emission of different hydrogen populations and determine properties of the heliospheric interface, measurements of Lyman-$\alpha$ spectra are required.

A spectrum of backscattered Lyman-$\alpha$ radiation is determined by the Doppler effect and depends on the projection of the distribution function of H atoms on the LOS. Thus, the analysis of the spectral properties of backscattered Lyman-$\alpha$ radiation makes it possible to obtain strong constraints on the hydrogen distribution function inside the heliosphere and derive information on the processes occurring in the vicinity of the heliopause.

\begin{figure*}
\centering
\includegraphics[width=1.0\textwidth]{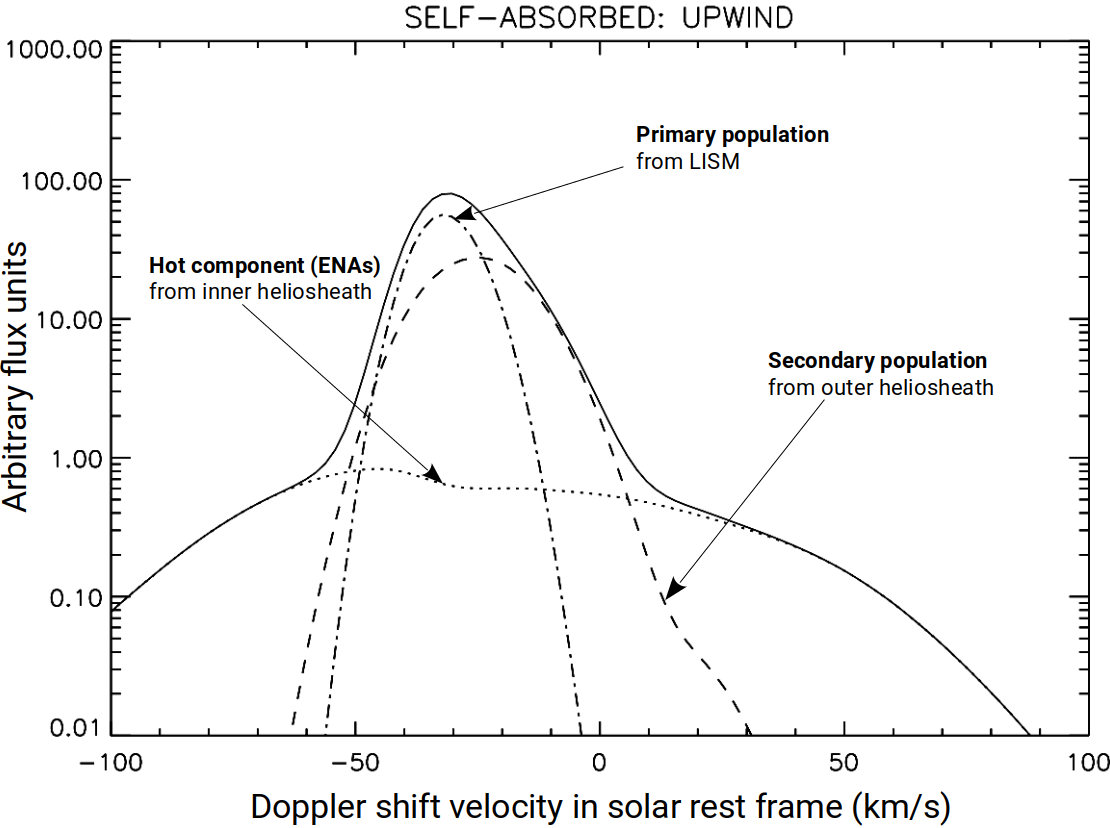}
\caption{
Modeled interplanetary Lyman-$\alpha$ line profile in the upwind direction. The simulation was performed in the self-absorption approximation for an observer at 1 AU in the upwind direction from the Sun. The Doppler shift velocity in the solar rest frame is presented on the x-axis in km/s. The emission due to the primary H population is shown by the dash-dotted line, the secondary hydrogen -- by the dashed line, the hot component -- by the dotted line.  The solid line presents the total emission due to the three hydrogen populations. Adapted from \citet{quemerais_izmod_2002}.
}
\label{fig:model_line_profile}
\end{figure*}

The influence of the heliospheric boundary on the backscattered Lyman-$\alpha$ emission profiles was studied by \citet{quemerais_izmod_2002} based on a self-consistent model of the SW/LISM interaction \citep{izmodenovetal2001}. Figure \ref{fig:model_line_profile} presents a modeled Lyman-$\alpha$ line profile in the upwind direction, with the superposition of three populations of H atoms in the heliosphere \citep[see also][]{izmodenovetal2001}:
\begin{enumerate}
    \item The primary population of the H atoms penetrates the heliosphere directly from the LISM. The contribution of this population to the total line is shown by the dash-dotted line.
    Note that this population is accelerated and cooled due to the selection effect (the slower the atom, the higher probability to be ionized). 
    \item The secondary H atoms are slower and hotter than primary because they are created by the charge exchange of protons in the outer heliosheath, where the interstellar plasma is decelerated and heated, with primary H atoms. The emission produced by the secondary population is presented by the dashed line in Figure \ref{fig:model_line_profile}, and it has a smaller Doppler shift compared to the LISM population line. 
    \item The third population, which we refer to as the hot component, is created by charge exchange of the core solar wind protons and pickup protons in the inner heliosheath (so-called energetic neutral atoms, ENAs), where the solar wind is decelerated to subsonic speed and the local plasma temperature of the order 10$^5$ -- 10$^6$ K \citep[see also][]{sokol_sameissue, galli_sameissue}. It is seen in Figure \ref{fig:model_line_profile}, that the hot component produces a very broad spectrum, and its peak emission is two orders of magnitude lower than the peak value of the total line. \citet{quemerais_izmod_2002} estimated that the contribution of the hot component to the backscattered emission intensity in the upwind direction is less than 5\% of the total intensity, and for the downwind direction it reaches 15\%. The spectrum of the hot component might be even broader (than it is shown in Figure \ref{fig:model_line_profile}) since the ENA population was simulated under the assumption of a Maxwellian distribution of protons in the inner heliosheath (which are parental to ENAs), while the pickup protons should be treated kinetically and separately from the core SW protons like it was performed by \citet{malama2006, chalov2015, baliukin2020, baliukin2022}. The observations of the Lyman-$\alpha$ emission from the hot component can set strong constraints on the properties of the inner heliosheath. Nevertheless, it requires a very high signal-to-noise ratio, which is not achievable by existing UV spectrographs.
\end{enumerate}
Also, a fourth component of hydrogen exists in the heliosphere, which is formed in the process of charge exchange of SW protons with H atoms in the supersonic solar wind region. However, this component is minor by its abundance, and the velocity of such atoms is $\sim$400 km/s (SW speed), which is out of the exciting solar Lyman-$\alpha$ line.

To distinguish contributions to the Lyman-$\alpha$ emission line of different hydrogen populations (see Figure \ref{fig:model_line_profile}), created in the heliosheath, from the hydrogen wall, or coming from the pristine LISM, as well as the Galactic emission component, a Doppler velocity resolution on the order of $\sim$5–10 km/s is required. This corresponds to a wavelength resolution of 0.002 – 0.004 nm or resolving power R $\sim$ 30,000 – 60,000. 
The highest resolution observations of the interplanetary hydrogen were performed by STIS on board HST (see Section \ref{sec:hst}) and Imaging Ultraviolet Spectrograph \citep[IUVS;][]{mcclintock2015, mayyasi2017} on board Mars Atmosphere and Volatile EvolutioN (MAVEN). However, the resolving power of these observations was only R $\sim$ 15,000 – 20,000, which corresponds to the velocity resolution of $\sim$15–20 km/s.

The other possibility to study the line profile is to use a hydrogen absorption cell, which contains $\rm H_2$ gas that is transparent to the Lyman-$\alpha$ photons. In the active state of the cell, the diatomic hydrogen gas is dissociated into monoatomic (utilizing the current passing through the filament), which is already not transparent and absorbs Lyman-$\alpha$ photons. Therefore, the cell can be used as a negative filter since it absorbs some fraction of the Lyman-$\alpha$ photons (depending on the Doppler shift between the cell and the incoming photon). The ratio of the intensity observations made with the hydrogen cell in the active and inactive states characterize the properties of the measured Lyman-$\alpha$ line profile. The hydrogen cell was first used in the 1980s on the soviet Prognoz-5/6 spacecraft, and the SWAN/SOHO instrument is the latest instrument to use this technique (see Section \ref{swan_hcell_data}).

\subsection{Early results: Mars-7 and Prognoz-5/6}

\citet{bertaux1976} analyzed the observations of Lyman-$\alpha$ emission on the Mars-7 spacecraft collected in 1973--1974. The spacecraft was equipped with a photometer and a hydrogen cell. As a result of the analysis, the first estimates for the average velocity and temperature of hydrogen gas in the heliosphere have been obtained: $T_{\rm H} = 12000 \pm 1000$ K, $V_{\rm H} = 19.5 \pm 1.5$ km/s. Note that further measurements showed that these estimates describe the parameters of interstellar H atoms after they pass through 
the termination shock (i.e. at distances of about 90 AU from the Sun) quite well.

The Prognoz-5 and 6 spacecraft performed measurements in 1976 -- 1978 on the high Earth orbit, which made it possible to minimize the influence of the Earth's geocorona. The spacecraft were oriented to the Sun and scanned the sky in a plane perpendicular to the direction of the Sun along a great circle. \citet{bertaux1977} provided the estimate for the LISM temperature ($T_{\rm LISM} = 8800 \pm 1000$~K) based on the first results of measurements by the Prognoz-5 spacecraft. Later, a series of papers by \citet{bertaux1984, lallement1984, lallement1985a, lallement1985b}, and \citet{bertaux1985}, was devoted to the analysis of the observations by the Prognoz-5 and 6 spacecraft. These studies were performed based on the classical hot model for the distribution of interstellar hydrogen in the heliosphere. The measurements using a hydrogen cell made it possible to determine the line width of backscattered Lyman-$\alpha$ radiation and the Doppler shift relative to the line center. The operational principle of the hydrogen cell and the method of data analysis are described in \citet{bertaux1984}. As a result of the analysis, the velocity vector of interstellar H atoms in the heliosphere was determined ($V_{\rm H} = 20 \pm 1$ km/s, $\lambda_{\rm H} = 71^{\circ} \pm 2^{\circ}$ -- ecliptic longitude of the vector $\textbf{V}_{\rm H}$, $\beta_{\rm H} = -7.5^{\circ} \pm 3^{\circ}$ -- ecliptic latitude of $\textbf{V}_{\rm H}$), the temperature of the LISM is $T_{\rm LISM} = 8000 \pm 1000$ K, and the number density of H atoms in the LISM is $n_{\rm H,LISM}=0.03 - 0.06$ cm$^{-3}$. In addition, it was shown that the intensity of the Lyman-$\alpha$ emission of a Galactic origin does not exceed 15 R, which is less than 5\% of the total intensity of backscattered emission measured in the heliosphere. Based on the hot model of the distribution of H atoms, \citet{lallement1985b} showed that Lyman-$\alpha$ emission data obtained in 1976 -- 1977 by Prognoz-5/6 can be explained by a 40\% decrease in the charge exchange ionization rate at the poles compared to the plane of the solar equator.

\subsection{HST/GHRS and STIS}\label{sec:hst}	

\begin{figure*}
\centering
\includegraphics[width=1.0\textwidth]{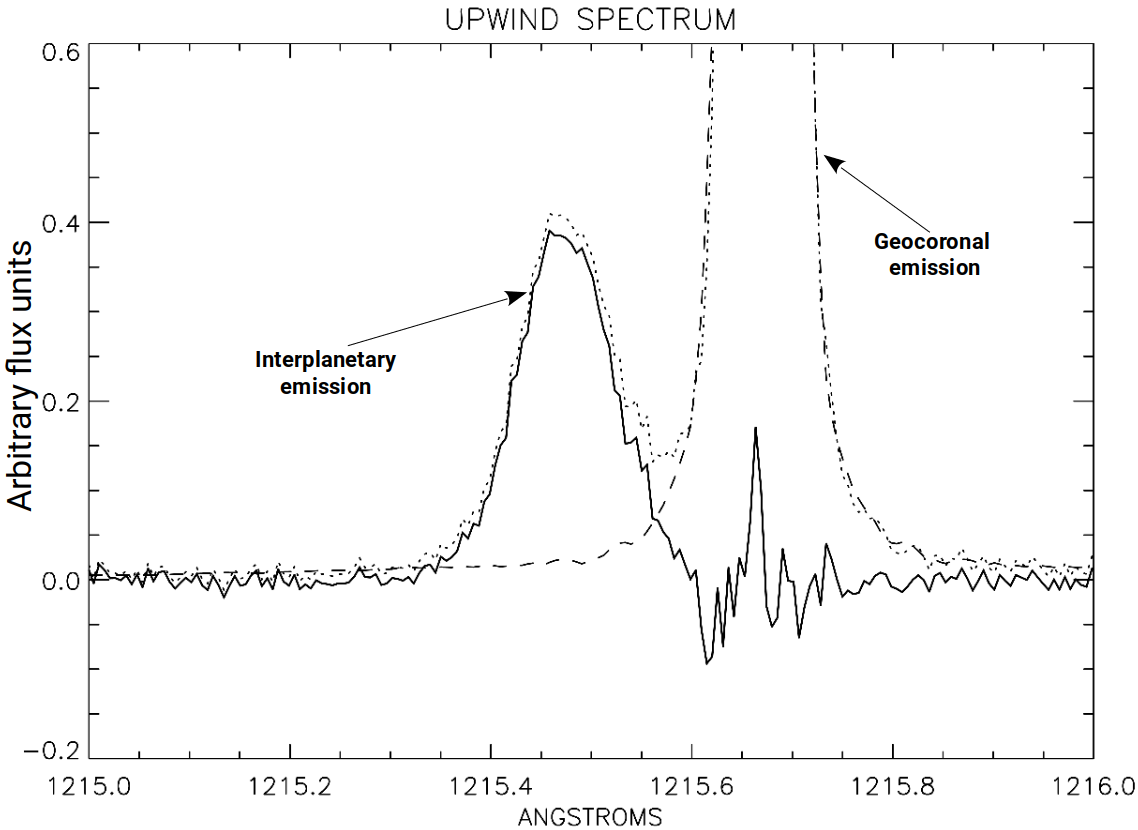}
\caption{
The Lyman-$\alpha$ line profile in the upwind direction observed by HST/STIS in March 2001. The dotted line presents the original profile. The geocoronal emission (dashed line) was removed from the data, and the rest is associated with the IP background emission (solid line). The x-axis shows the wavelength in the Earth rest frame in units of Angstrom, so the geocoronal line is centered at 1215.67 Å. Adapted from \citet{quemerais2009}.
}
\label{fig:HST_line_profile}
\end{figure*}

The interplanetary Lyman-$\alpha$ line profile have been observed by the Goddard High Resolution Spectrograph (GHRS) and Space Telescope Imaging Spectrograph (STIS), which replaced the GHRS in 1997, on board the Hubble Space Telescope \citep{clarke1995, clarke1998}.

Observations of HST are complicated by the Earth geocorona emission at Lyman-$\alpha$ (see also Section \ref{sec:geocorona}) seen from the low Earth orbit ($\sim$540 km). The emission is very bright and tends to mask the interplanetary Lyman-$\alpha$ line. The best time for measurements is when the Doppler shift between the Earth emission and the interplanetary line is largest, which occurs in March each year. Figure \ref{fig:HST_line_profile} shows an example measurement of the upwind IP line profile obtained by STIS on board Hubble Space Telescope in early March 2001. As can be seen from these data, it is impossible to identify the hot component of H atoms (ENAs) that originated in the inner heliosheath because the signal-to-noise ratio is insufficient.

\citet{linsky_wood1996} have analyzed the Lyman-$\alpha$ absorption spectra recorded by HST/GHRS  toward $\alpha$ Centauri and discovered the solar system hydrogen wall, a pile-up of the H number density in front of the heliopause. In addition to the ISM and heliospheric absorptions, the Lyman-$\alpha$ absorption spectra towards nearby stars illustrate absorption from the analogous "astrosphere" surrounding the star \citep{wood2005}. Since the H wall neutrals are heated and decelerated, for an observer inside the heliosphere the heliospheric absorption is redshifted (relative to the ISM absorption), while the analogous astrospheric absorption is blueshifted, so these absorptions avoid complete obscuration by the ISM.

The interplanetary hydrogen (IPH) line shift (or apparent velocity) has been estimated from GHRS and STIS data by several authors \citep{clarke1998, scherer1999, benjaffel2000, quemerais2006c}. One of the most recent studies by \citet{vincent2011} presented an updated analysis of IPH velocity measurements from GHRS and STIS during solar cycle 23 and summarized all previous estimations of the line shift \cite[compiled in tables 2 and 3 of][]{vincent2011}.

To observe the IPH along a line of sight, STIS has been used in cross-dispersed echelle mode with the long slit ($\SI{52}{\arcsecond} \times \SI{0.5}{\arcsecond}$). Using a long slit allows the collection of more photons and increases the signal-to-noise from an extended source (such as IPH). However, this configuration results in the superposition of different orders of the echelle spectrum and contamination of the Lyman-$\alpha$ line (121.6 nm) by overlapping light from the 130.4 nm triplet line of geocoronal oxygen ($\rm O I$) in lower orders.
\citet{vincent2011} showed a significant change in apparent velocity due to this contamination, not accounted for in earlier reports.
Their reanalysis of the observations of HST/GHRS and HST/STIS provided IPH line shifts of 22.2 $\pm$ 1.5, 24.0 $\pm$ 0.9, and 22.4 $\pm$ 0.4 km/s in 1994, 1995, and 2001, respectively, consistent (within 1$\sigma$) with predictions of the two different numerical models by \citet{quemerais2008} and \citet{scherer1999}. However, some discrepancy has been found in comparison with SOHO/SWAN observations in 1997 and 1998.

\citet{vincent2014} studied the IPH measurements provided by the HST/STIS during solar cycle 24. The results of the analysis have been compared with SWAN observations and simulations of the global heliospheric model by \citet{izmod2013} using two sets of LISM physical parameters based on Ulysses \citep{witte2004} and IBEX \citep{mccomas2012} measurements, which predict different LISM velocity vectors. The authors showed that the IPH data (both from STIS and SWAN) fit Ulysses-based model results better than IBEX-based model results. Their analysis of HST/STIS observations also provided IPH line shifts of 23.5 $\pm$ 0.5 km/s and 23.3 $\pm$ 0.5 km/s in 2012 and 2013, respectively.

\subsection{SOHO/SWAN hydrogen cell observations} \label{swan_hcell_data}

The absorption hydrogen cell of the SWAN instrument is placed in the optical path and filled with $\rm H_2$ gas, which is transparent to the Lyman-$\alpha$ emission. When the cell is activated, two tungsten filaments become heated. They dissociate the molecules into atoms, which results in a cloud of H atoms that scatters the Lyman-$\alpha$ photons near the line center. The characteristics of the hydrogen cell are the optical thickness at line center $\tau_{\rm c}$ and temperature $T_{\rm c}$, and for the SWAN instrument, these parameters are $\tau_{\rm c} \approx 3$ and $T_{\rm c} \approx 300$ K \citep[see][]{quemerais1999}. For each LOS, the photons are counted first with the cell switched off, giving the intensity $I_{\rm off}$, and then with the H cell switched on, which gives the intensity $I_{\rm on}$. 

\begin{figure*}
\centering
\includegraphics[width=1.0\textwidth]{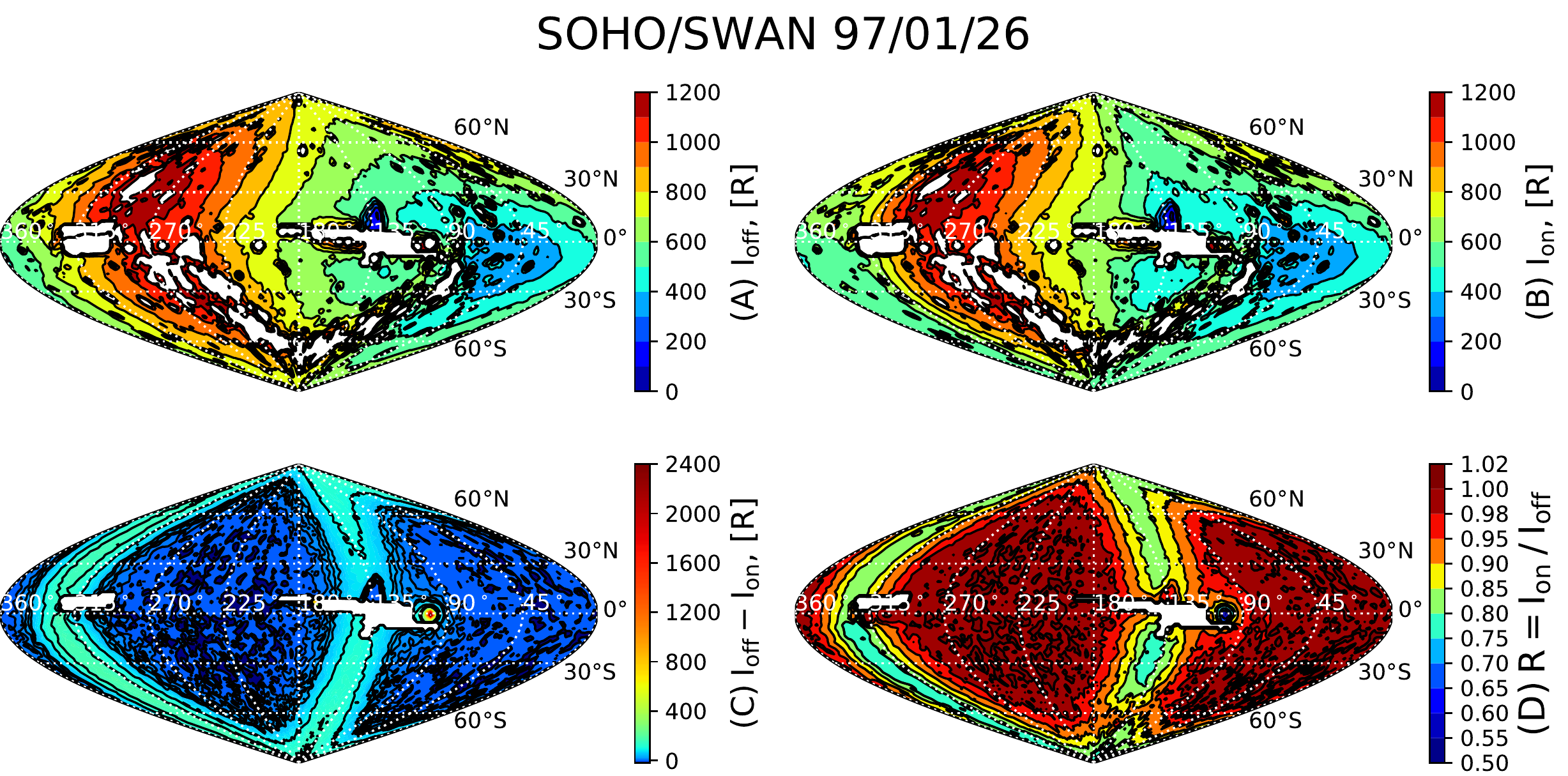}
\caption{Full sky sinusoidal projections (in ecliptic J2000 coordinates) of the Lyman-$\alpha$ intensities $I_{\rm off}$ (panel A), $I_{\rm on}$ (panel B), $I_{\rm off} - I_{\rm on}$ (panel C), and reduction factor $R = I_{\rm on} / I_{\rm off}$ (panel D) observed by SWAN/SOHO on 26 January 1997. Adapted from \citet{baliukin2019}.
}
\label{fig:swan_fullsky_2}
\end{figure*}

Figure \ref{fig:swan_fullsky_2} presents the distribution (in ecliptic coordinates) of Lyman-$\alpha$ intensities, which were measured by SWAN on 26 January 1997: panel A shows the intensity $I_{\rm off}$, panel B represents the intensity $I_{\rm on}$, panel C shows the difference of intensities observed at two hydrogen cell states ($I_{\rm off} - I_{\rm on}$), and panel D is the ratio $R = I_{\rm on} / I_{\rm off}$ (a so-called reduction factor). The great circle that passes through the ecliptic poles is seen in panels B and C, and it is called a Zero Doppler Shift Circle (ZDSC). This circle consists of directions perpendicular to the relative motion between the interstellar hydrogen flow and SOHO, so for these directions, the interplanetary hydrogen is absorbed by the H cell the most \citep{bertaux1984}. The motion of the ZDSC along the year includes the composition of the orbiting Earth’s velocity vector ($\sim$30 km/s) with the fixed vector of the interstellar flow velocity of about 26 km/s. At $\sim 100^\circ$ ecliptic longitude and zero latitude, there is a brighter region, which is produced by the Lyman-$\alpha$ emission from the geocorona, even more conspicuous on panel C (see Section \ref{sec:geocorona}).

\citet{costa1999} performed a parametric study based on a single-component hot model and determined the parameters of H atoms at 50 AU from the Sun. The results of \citet{costa1999} confirmed the existence of secondary interstellar H atoms, which leads to a decrease in the velocity and an increase in the temperature of the total flow of atoms at 50 AU from the Sun in comparison with the parameters of the primary interstellar atoms in the LISM. Using the hydrogen cell absorption map, \citet{costa1999} derived the LOS temperature (or linewidth) variation with the angle from the upwind direction directly from the data and found a temperature minimum between the upwind and crosswind directions, while classical models predict a monotonic increase of the linewidth from upwind to downwind. The existence of two distinct populations of H atoms (primary and secondary) with different velocities was suggested to explain this behavior.

\subsubsection{Line profile reconstruction}

\citet{quemerais1999} have developed a technique to derive the interplanetary Lyman-$\alpha$ line profiles by accumulating one year of hydrogen absorption cell measurements performed by the SWAN instrument on board SOHO. For a given fixed LOS, the spectral Doppler scan is provided by the orbital velocity of the Earth ($\sim$ 30 km/s) over one full year. The achieved Doppler scan is $\pm$ 30 $\times \cos \beta$ km/s, where $\beta$ is the ecliptic latitude of the LOS. The spectral scan across the Lyman-$\alpha$ line allowed deriving the line profiles and, consequently, the velocity distribution. This technique is independent of both the hydrogen distribution model and the multiple scattering effects. Thus, the results represent the ground truth, which puts strong constraints on the global heliospheric models of SW/LISM interaction.

\citet{quemerais2006c} continued the work on the line profile reconstruction technique. While \citet{quemerais1999} were focused only on the first year of measurements, \citet{quemerais2006c} have reconstructed the interplanetary Ly-$\alpha$ line profiles using the SWAN hydrogen cell observations from the solar minimum in 1996 to the solar maximum in 2001 and found that the line shift (LOS apparent velocity) of the interplanetary Lyman-$\alpha$ line in the upwind direction varies significantly (from 25.7 km/s to 21.4 km/s from solar minimum to solar maximum, respectively). The analysis of the line width (LOS temperature) suggested that the interplanetary line consists of two components, which are produced by the backscattered photons by two distinct populations of hydrogen that have different parameters (bulk velocities and temperatures). This circumstance is an obvious sign of the heliospheric interface influence on the line profiles observed at 1 AU from the Sun. The SWAN data comparison with an interplanetary background upwind spectrum obtained by HST/STIS in March 2001 yielded a good agreement (within the uncertainties and possible biases involved in the analysis).

\subsubsection{Estimation of the hydrogen flow direction}

Another extremely important result obtained using the SWAN data is the proof of the effect of the interstellar magnetic field (IsMF) on the heliosphere. A model-independent method of Lyman-$\alpha$ line profile reconstruction developed by \citet{quemerais1999} allowed also to estimate the H flow direction ($\lambda$, $\beta$)$_{\rm H}$ = ($252.3^\circ \pm 0.73^\circ$, $8.7^\circ \pm 0.9^\circ$) in J2000 ecliptic coordinates. 

Later, \citet{lallement2005, lallement2010} analyzed SWAN data obtained with the hydrogen cell using a single-component hot model of the hydrogen distribution in the heliosphere. Temperature and bulk velocity of atoms at 80 AU from the Sun in the model were set based on previous results from \citet{costa1999}, and the bulk velocity direction was varied. As a result, the direction of the velocity in the heliosphere, which leads to the best agreement between theory and SWAN measurements, was determined. The values (ecliptic J2000 coordinates) were found with two independent analyses ($\lambda$, $\beta$)$_{\rm H}$ = (252.5$^\circ$ $\pm$ 0.7$^\circ$, 8.9$^\circ$ $\pm$ 0.5$^\circ$), which differs by $\sim$4$^{\circ}$ from the direction of motion of interstellar helium atoms in the LISM, i.e. ($\lambda$, $\beta$)$_{\rm He}$ = (255.4$^\circ$ $\pm$ 0.5$^\circ$, 5.2$^\circ$ $\pm$ 0.2$^\circ$) \citep{witte2004}.

The plane formed by velocity vectors of H and He atoms is commonly called the Hydrogen Deflection Plane (HDP). It should be clarified here that, unlike H atoms, interstellar helium atoms penetrate the heliosphere almost freely since they do not interact with protons (due to the small cross-section for charge exchange and large ionization potential). Therefore, it is believed that the direction of motion of helium atoms in the heliosphere, which is known from both direct measurements of helium by the GAS instrument on board the Ulysses spacecraft and indirect measurements of the UV radiation of the Sun backscattered by them, coincides with the direction of the incident flow of the LISM. Thus, \citet{lallement2005, lallement2010} showed that during the passage of the region of the heliospheric interface, the direction of the average velocity of H atoms deviates by several degrees compared to the direction of their velocity in the LISM. This effect was explained by the influence of the interstellar magnetic field. The point is that if the interstellar magnetic field is inclined with respect to the LISM velocity vector relative to the Sun, then the LISM plasma flow in the vicinity of the heliopause will become asymmetric \citep{izmod2005b}. Therefore, secondary interstellar H atoms produced as a result of charge exchange on protons near the heliopause will, on average, be deviated from the initial direction of the LISM motion. This is exactly what is observed in the SWAN data on the backscattered solar Lyman-$\alpha$ radiation. The effect of hydrogen deflection due to the influence of the interstellar magnetic field was soon confirmed theoretically by \citet{izmod2005b} based on the analysis of the simulation results of the global self-consistent kinetic-MHD model of the heliospheric interface, which take into account the interstellar magnetic field.

However, the direction of interstellar hydrogen flow in the vicinity of the Sun may be influenced by other effects such as the solar radiation pressure and gravitation, ionization, and kinetic non-Maxwellian properties of the hydrogen distribution and also may depend on other LISM parameters besides IsMF. \citet{katushkina2015b} have performed theoretical modelling of the backscattered solar Lyman-$\alpha$ radiation seen at 1 AU from the Sun and analyzed the direction of H flow in the heliosphere, which can be obtained from the spectral properties of the backscattered radiation, and investigated the influence of different effects mentioned above. The results of simulations in the frame of the state-of-art 3D time-dependent kinetic model of the H distribution, which is based on the global heliospheric model by \citet{izmod2015}, were compared with the SWAN data of 1996. \citet{katushkina2015b} concluded that deflection of the interstellar hydrogen flow in the heliosphere is potentially a very powerful tool to estimate the IsMF, but in doing this the considered effects need to be taken into account.

\citet{koutroumpa2017} suggested a method to determine the longitude of the interstellar H flow, which is model-independent and based on the parallax effect provided by the SWAN motion around the Sun. Even without the use of the H cell data, \citet{koutroumpa2017} have shown that for 20 years covered by the SWAN dataset, the longitude of the interstellar hydrogen flow vector varies insignificantly from its average value of $252.9^\circ \pm 1.4^\circ$, which reinforces the assumption of the interstellar gas flow stability.

\subsubsection{Geocorona} \label{sec:geocorona}

\citet{baliukin2019} have studied the Lyman-$\alpha$ observations of the geocorona performed by the SOHO/SWAN in January 1996, 1997, and 1998 (low solar activity). The use of the H cell allowed assigning almost the entire difference of intensities $I_{\rm off}-I_{\rm on}$ to the geocorona emission since, in this region of the sky, the interplanetary line is Doppler shifted with respect to the H cell. Still, the small contribution from IP background absorption was simulated and subtracted from the original data. To calculate the IP background, the model by \citet{katushkina2015a} with the boundary condition at 70 AU from the global heliospheric simulations by \citet{izmod2015} was used. The major result is that the emission from the geocorona was found to extend up to $\sim$100 $R_{\rm E}$ from Earth ($6.4 \times 10^5$ km), which is far beyond the Moon's orbit ($\sim$ 60 $R_{\rm E}$) and exceeding the earlier estimate ($\sim$ 50 $R_{\rm E}$) obtained with the LAICA imager \citep{kameda2017}. For a distant observer outside the exosphere, the geocoronal intensity is of the order of 5 R at $\sim$100 $R_{\rm E}$ and about 20 R at the Moon's orbit. The geocorona may be an unwanted source of an excess Lyman-$\alpha$ emission for an observatory dedicated to the universe exploration in UV.

\section{Conclusions}

In this paper, we summarized a knowledge about the heliosphere deduced from numerous measurements (both photometric and spectrometric) of the backscattered solar Lyman-$\alpha$ emission collected since the beginning of the space era. We illustrated that analysis of Lyman-$\alpha$ observations using physics-based global models has proven itself as an effective tool for indirect diagnostics of the parameters of H atoms in the heliosphere, properties of the interstellar medium, and the physical processes at the heliospheric boundary. The following summarizes major discoveries enabled by the studies and analysis of the backscattered solar Lyman-$\alpha$ emission:
\begin{itemize}

\item[(i)] Backscattered Lyman-$\alpha$ observations proved that parameters of interstellar hydrogen in the solar system are different from those of interstellar helium. Indeed, the interstellar hydrogen is decelerated by $\sim3 - 4$ km/s and heated by $\sim$5000 K compared to helium. This provides observational evidence for the effective charge exchange process between interstellar hydrogen and slowed down and heated interstellar plasma around the heliosphere. Similarly, such interaction would create slower and warmer interstellar hydrogen around other astrospheres. Hence, the Lyman-$\alpha$ observations confirm an existence of hydrogen wall.

\item[(ii)] SWAN/SOHO Lyman-$\alpha$ observations with the H cell discovered the $4^\circ$ deflection of the hydrogen flow in the heliosphere relative to the pristine flow in the LISM. This is a signature of an asymmetry of the global heliosphere caused by the interstellar magnetic field.

\item[(iii)] Analysis of the SOHO/SWAN intensity maps established variations of the solar wind mass flux with heliolatitude and time over more than two solar cycles. In particular, two maxima at mid-latitudes were discovered during the solar activity maximum (of the solar cycle 23), which Ulysses missed due to its specific trajectory.

\item[(iv)] Lyman-$\alpha$ observations in the outer heliosphere on Voyager/UVS and New Horizons/Alice revealed a presence of the extraheliospheric emission with a magnitude in the order of a few tens Rayleigh.

\end{itemize}
These discoveries were key to advance our understanding of the ion-neutral coupling processes determining the dynamic interaction of the heliosphere with the LISM. However, a number of open questions exist that have to be addressed with the future Lyman-$\alpha$ experiments and studies.

\subsection{Open questions and future developments}
We outline a list of investigations which can be done in the nearest (or not in so nearest) future to advance the field and obtain more valuable information from the existing and future Lyman-$\alpha$ data. Investigations can be divided into two categories depending on a type of measurements, photometric or spectrometric.

The following lists investigations with photometric Lyman-$\alpha$ observations (intensity):
\begin{enumerate}
    \item
    Various studies of Lyman-$\alpha$ emission \citep[e.g.,][]{quemerais1996, pryor2008} determined the reference number density of interstellar hydrogen in the range of $\sim 0.08-0.15$ cm$^{-3}$. These values refer to the density at large distances (e.g., at 70--80 AU) in the upwind direction. Two factors contribute to a large uncertainty: (1) the uncertainties and issues in the calibration of Lyman-$\alpha$ instruments (the major effect), and (2) the differences in methodology to determine the number density from the Lyman-$\alpha$ data obtained mostly at a small heliocentric distance (1 AU). 
    
    The issues and details on calibration and inter-calibration have been reported by \citet{quemerais2013book}. There are ways to improve and verify existing calibration using numerical models. For instance, a comparison of long-term observations with results of model calculations of Lyman-$\alpha$ intensities allows estimation of the sensitivity degradation of a particular instrument. This approach was applied to the SOHO/SWAN 1996--2018 dataset. A preliminary result shows that the sensitivity of the SWAN sensor units degrades at a rate of $\sim$2\% per year. This should be taken into account in the future data analysis.
    
    The second factor is connected with the fact that in most studies, a hot model and its modifications are applied to obtain the reference number density. These models assume that parameters of interstellar hydrogen are uniform at large heliocentric distances ($\sim$70-80 AU) and velocity distribution is Maxwellian or bi-Maxwellian. These assumptions are not confirmed by self-consistent models of the global heliosphere that show both radial and angular gradients in the number density. A straightforward way is to use a global model of the heliosphere and to use the H atom number density in the undisturbed LISM as the reference number density. Global models of the heliosphere require values of interstellar proton number density and magnitude and direction of the undisturbed interstellar magnetic field. These values can be derived using different observations (for instance, from Voyager) and multi-parametric fitting of the model results with various data. \citet{izmod2003} performed such study and estimated number densities of H atoms and protons in the undisturbed LISM: $n_{\rm H,LISM} = $ 0.2 cm$^{-3}$ and $n_{\rm p,LISM} = $ 0.04 cm$^{-3}$, respectively. Recently, \citet{bzowski_etal:19a} performed a parametric analysis to determine the ionization state of the LISM and concluded that $n_{\rm H,LISM} = $ 0.154 cm$^{-3}$ and $n_{\rm p,LISM} = $ 0.054 cm$^{-3}$. Both \citet{izmod2003} and \citet{bzowski_etal:19a} used the reference number density of interstellar hydrogen in the outer heliosphere obtained by \cite{bzowski_etal:08a} in the analysis of pickup ion observations from Ulysses \citep{gloeckler_etal:92, gloeckler_geiss:01a}, and the slowdown of solar wind due to mass-loading by pickup ions, measured by Voyager \citep{richardson_etal:08a}.
    
    However, a recent analysis of the pickup ion data from New Horizons/SWAP \citep{swaczyna_etal:20a} suggested that the density of ISN H at the termination shock is larger by almost 40\% than previous estimates. Additionally, \citet{swaczyna_etal:21a} showed that elastic collisions in the outer heliosheath heat ISN H and He by $\sim$1000 K. Thus, LISM temperature determined from in-situ measurements on board Ulysses and from IBEX is overestimated. The actual temperature of the LISM could be about 6500 K (from IBEX) or about 5300 K (from GAS). However, existing models of the heliosphere, to our knowledge, do not take into account these heating effects.
    
    An updated analysis of newly calibrated Lyman-$\alpha$ intensity data must be performed to derive independently both the reference number density of interstellar hydrogen in the outer heliosphere and the LISM parameters. The latter can be done using new state-of-the-art models which include recently recognized physical processes such as effects of elastic collisions or the electron thermal conduction in the inner heliosheath \citep{izmod2014}.

    \item Another direction of future research is related to deriving the solar wind flux as a function of the heliolatitude and time. 
    For the first time, a possibility of using measurements of the backscattered solar Lyman-$\alpha$ emission for diagnostics of the inhomogeneous structure of the solar wind was noted by \citet{joselyn1975}. \citet{quemerais2006b, lallement2010} applied inversion methods to the SOHO/SWAN Lyman-$\alpha$ maps.
    On the one hand, recent analysis of SOHO/SWAN sky mapping data \citep{katushkina2013, katushkina2019, koutroumpa2019} suggest that solar wind flux maxima at mid-latitudes occur at maxima of solar activity. WSA-Enlil model of the solar wind propagation show a qualitative agreement with the derived structure.
    On the other hand, analysis of interplanetary scintillation (IPS) data \citep{sokol_etal:13a, bzowski_etal:13b, sokol_etal:20a, porowski_etal:21a} shows that these maxima may occur from time to time but do not seem to be a regular solar wind feature.
    However, the IPS observations \citep{tokumaru2013, tokumaru2021} provide only heliolatitudinal profiles of solar wind speed on a yearly timescale. To compute the flux, the SW density profile is required. To calculate the density, \citet{sokol_etal:13a, bzowski_etal:13b} used the SW velocity-density relation derived from observations by Ulysses. For the same purpose, \citet{sokol_etal:20a, porowski_etal:21a} used the empirically derived (based on HELIOS, Wind, and Ulysses data) latitudinal invariance of the SW energy flux \citep{lechat2012}. However, the assumption about the invariance is not physically justified.
    
    Will the solar flux structure have maxima at mid-latitudes in the upcoming solar maximum (solar cycle 25)? What are the physical processes responsible for their formation? These questions remain to be answered.
   
    \item Interstellar H atoms surrounding the Sun behaves like a screen on which solar Lyman-$\alpha$ photons are projected. Active regions on the Sun (solar flares) produce higher intensities of the backscattered Lyman-$\alpha$ emission, and they are visible in the sky maps. As shown by \citet{bertaux2000} and \citet{quemerais2002}, Lyman-$\alpha$ intensity sky maps can be utilized to forecast the solar activity in advance up to two weeks (a half the solar rotation period). Additionally, by applying an inversion technique to the maps, location and strength of the active regions on the far side of the Sun can be estimated. Future developments and automation of inversion algorithms will provide an important contribution to a wide range of tools used for space weather prediction.
    
\end{enumerate}

The following lists investigations with spectrometric Lyman-$\alpha$ observations (line profiles):

\begin{enumerate}

    \item A measurement of the deflection of the interstellar hydrogen flow in the heliosphere is potentially a powerful tool to constrain a direction of the magnetic field. \citet{lallement2005, lallement2010} performed analysis of the SOHO/SWAN data with H cell using a hot model of interstellar hydrogen with a single Maxwellian boundary condition. They determined a deflection of the hydrogen flow by $\sim 4^\circ$. However, later \citet{katushkina2015b} demonstrated that a two-component model is required to estimate the deflection. Additionally, one needs to take into account the solar radiation pressure, gravitation, ionization, and kinetic non-Maxwellian properties of the hydrogen distribution. Therefore, the SWAN reduction factor data should be re-analyzed using the two-component hydrogen model to determine the deflection and compare with previous finding.
    
    From the observational point of view, to differentiate contributions of primary and secondary populations of interstellar H atoms to the observed Lyman-$\alpha$ emission spectra, a very high spectral resolution ($\sim$5 km/s) is required. To investigate the structure and properties of the hydrogen wall, spectral observations from observing points far away from the Sun are required. At large distances from the Sun, the emission from the H wall becomes significant compared to the rapidly decreasing heliospheric emission.

    \item The hot component of H atoms which originates in the inner heliosheath (ENAs) has not been detected yet in Lyman-$\alpha$ observations. To do this, measurements need to have an extremely high signal-to-noise ratio, not achieved by the existing Lyman-$\alpha$ instruments. Analysis of the ENA distribution using Lyman-$\alpha$ spectra can serve as a new tool for studying a nature of the heliosheath. A feasibility study of using Lyman-$\alpha$ spectra to constrain ENA distribution will be valuable in light of the planned future space missions (such as Interstellar Probe, see Section \ref{sec:isp}), which carries a UV spectrometer in an example payload. From the modelling perspective, to make a correct estimate of the hot component contribution to the observed intensities of the interplanetary Lyman-$\alpha$ emission, the ENAs and their parental component (pickup protons) should be simulated with a kinetic approach.
    
    \item SOHO/SWAN data show a minimum of the Lyman-$\alpha$ linewidth near the crosswind direction, while all numerical models with a single component of H atoms predict its monotonic increase from upwind to downwind. \citet{costa1999} and \citet{quemerais2006c} suggested that the existence of primary and secondary hydrogen atoms can explain this behavior.
    \citet{katushkina2011} have explored a nature of the linewidth minimum theoretically and proven this assumption using model simulations. However, the minimum of the Lyman-$\alpha$ linewidth appears only for the two-component hot model, as shown by \citet{katushkina2011}, while there is no distinct minimum in the simulations with the model that takes into account an influence of the heliospheric interface. It means that the non-Maxwellian features of the velocity distribution of H atoms (such as strong anisotropy of the kinetic temperatures) caused by the heliospheric interface compensate for the effect of two populations. This explanation is also supported by the recent calculations based on the state-of-the-art global heliospheric model by \citet{izmod2020}. 
    
    Therefore, the nature of the Lyman-$\alpha$ linewidth minimum in the data remains unknown.
    
    One can assume that the minimum can be explained by the presence of the hot population of H atoms (ENAs) in the heliosphere, but numerical calculations to justify/reject this assumption have not been performed yet. More detailed spectroscopic measurements and/or numerical simulations are needed to resolve this question.
    
    \item None of the existing models of hydrogen distribution in the heliosphere is capable to quantitatively reproduce the observed by SWAN/SOHO Lyman-$\alpha$ intensities from the upwind and downwind directions simultaneously. What is the reason for the discrepancy between models and data, and what physical processes are missing in current models? 
    
    A possible answer is the inclusion of the absorption of solar Lyman-$\alpha$ photons by H atoms in the model of the solar radiation pressure, which may explain the differences at least partially. \citet{kowalska2022} showed that the effect of absorption modifies the hydrogen distribution through the reduction of the radiation pressure force and, as a consequence, it generally provides higher (by a few percent) number density, especially in the downwind direction. However, an impact of the modified hydrogen distribution on the simulated Lyman-$\alpha$ intensities was not studied yet.
    
    The other aspect missing in most models is the hot component of H atoms. \citet{quemerais_izmod_2002} performed calculations of the backscattered Lyman-$\alpha$ emission produced by the hot component. They found that its contribution to the total intensity seen from 1 AU can reach 15\% in the downwind direction. However, this simulation was based on the assumption of a Maxwellian distribution of protons, but the population of pickup protons should be treated kinetically to estimate the actual contribution of the hot component.
 
    \item 
    \begin{table*}
    \caption{Estimates of the extraheliospheric Lyman-$\alpha$ component emission.}
    \label{tab:extraheliospheric}
    \begin{tabular}{lllll}
    \hline\noalign{\smallskip}
    Reference                   & Data source           & Observation   & Distance to   & Estimation \\
                                &                       & years         & the Sun (AU)  & (Rayleigh) \\
    \noalign{\smallskip}\hline\noalign{\smallskip}
    \citet{lallement1984}       & Prognoz-5 and 6       & 1976-1977     & 1             & $\leq 15$ \\
    \citet{lallement2011}       & Voyager 1(2)/UVS      & 1993–2003     & 53–88 (38–70) & 3–4 \\
    \citet{katushkina2017}      & Voyager 1/UVS         & 2003–2014     & 90–130        & 25 \\
    \citet{gladstone2018}       & New Horizons/Alice    & 2007–2017     & 8–40          & 40 \\
    \citet{strumik_etal:20a}    & SOHO/SWAN, IUE        & 1997          & 1             & 28.9$^a$ \\
    \citet{gladstone2021}       & New Horizons/Alice    & 2015–2020     & 32–47         & 43 $\pm$ 3 \\
    \noalign{\smallskip}\hline \\
    \end{tabular}
    \\{\bf Note.} {\footnotesize$^a$Background from unresolved point sources that was evaluated specifically for observations of the SWAN instrument.}
    \end{table*}
    
    Several authors, mentioned earlier in this chapter, reported evidence of an excess emission on top of the helioglow (see Table \ref{tab:extraheliospheric}, which summarizes existing estimations of the extraheliospheric background). The question of the origin (Galactic, extragalactic, contamination by UV continuum of stars, resolved or unresolved) and spatial distribution of the extraheliospheric emission is connected with the exploration of the interstellar matter in the Galaxy and beyond and can be resolved using spectrometric measurements. The extraheliospheric component of Lyman-$\alpha$ emission can be studied directly far away from the Sun, where it is not obscured by the interplanetary glow. It is important to note that the extraheliospheric background is neglected in most analyses of the helioglow, but its consideration could probably affect the results. 
  
    
\end{enumerate}
These questions will hopefully be addressed by the future missions, such as GLObal solar Wind Structure (GLOWS) experiment on board Interstellar Mapping and Acceleration Probe (IMAP; see the following Section \ref{sec:imap}) and a UV spectrometer on an Interstellar Probe (Section \ref{sec:isp}).

\subsection{Future missions}

\subsubsection{IMAP/GLOWS} \label{sec:imap}

The forthcoming NASA mission IMAP is scheduled for launch in 2025 into a Lissajous orbit around the Lagrange point L1 \citep{mccomas_etal:18b}. GLOWS is one of ten instruments on IMAP, and it will measure the heliospheric resonant backscattered Lyman-$\alpha$ emission. The science objective of GLOWS is to investigate a global heliolatitudinal structure of the solar wind and its evolution during a solar cycle. Additionally, GLOWS investigates the distribution of ISN hydrogen and the solar radiation pressure acting on ISN H. The GLOWS detector is a non-imaging single-pixel Lyman-$\alpha$ photometer, effectively a photon counting instrument. It is conceptually based on the Two Wide-angle Imaging Neutral-atom Spectrometers (TWINS)/LaD photometer \citep{nass_etal:06a}. The instrument includes a collimator with a baffle, a spectral filter, and a channeltron (CEM) detector. They are connected to an electronics block, responsible for collecting the event pulses and binning them for downlinking to the ground. The instrument is designed and assembled in the Space Research Centre PAS (CBK PAN) in Warsaw, Poland. 

The idea behind the observation strategy of GLOWS goes back to analysis of the evolution of the solar wind structure by \citet{bzowski:03, bzowski_etal:03a}. These authors took SWAN observations performed from the downwind and upwind locations of the SOHO spacecraft and limited to Sun-centered great circles in the sky. They assumed that the solar wind is bi-modal, slow and dense at the ecliptic plane and fast and rarefied in the polar regions. Also, they assumed that transitions from slow to fast wind occur rapidly as a function of heliolatitude, and the transition latitudes vary during a cycle of solar activity, independently in the north and south hemispheres. They allowed for hypothetical differences between the solar wind flux at the north and south poles and simulated the expected signal on a grid of the fast/slow boundaries and pole/ecliptic flux ratios. With this, they simulated the expected helioglow profiles along the selected great circles in the sky on a grid of contrast magnitudes and boundary latitudes. They determined the best-matching boundaries and contrast magnitudes between the north and south-polar flux and that in the ecliptic plane between the solar minimum in 1996 and solar maximum in 2002. As verified much later \citep{bzowski_etal:13a}, these results turned out to be similar to those obtained from analysis of interplanetary scintillations.

Based on this insight, GLOWS will scan a 75$^\circ$ circle in the sky, centered at a point offset by 4$^\circ$ from the Sun towards lower ecliptic longitudes. With this, the lines of sight will intersect the solar polar axis approximately within the maximum emissivity region, and therefore, the observations are expected to be the most sensitive to the solar wind imprints in the density distribution of ISN H gas \citep{kubiak_etal:21b}. 

The diameter of the field of view of the instrument is $\sim$4$^\circ$ full width at half maximum (FWHM). Such a narrow field of view, provided by a specially designed entrance system with a collimator, a baffle, and a sunshield, eliminates solar glint and any diffuse radiation from the outside of the desired region of the sky.  After a day of observations, the spin axis of IMAP is redirected to maintain the 4$^\circ$ offset from Sun's center, and the strip observed by GLOWS accordingly shifts in the sky.   

Since the extraheliospheric sources, both those bright and spatially resolved, and those dim and unresolved were found to provide a significant contribution to the SWAN signal \citep{strumik_etal:20a}, which is quite challenging to mitigate, GLOWS takes measures to eliminate them almost entirely. This is accomplished by using a narrow-band interference filter, which lets in only a waveband of $\sim$5 nm. 
While the helioglow emission is practically monochromatic and close to the Lyman-$\alpha$ wavelength, the stellar and galactic emissions are not. Applying a narrow-band filter suppresses the light from the outside of the filter transmission band. As a result, with the filter, the relative contribution of starlight in the total signal is less than without the filter because a large portion of starlight is cut out.
With this, the expected magnitude of the stellar contribution is lower from that found for SWAN by a factor of 20.

Since GLOWS has only one detector, challenging issues of intercalibration between two SWAN sensors observing the northern and southern hemispheres will be absent. The planned GLOWS observations will be binned at a high resolution (3600 bins for 360$^\circ$ of the scanning circle), so that contributions from bright stars can be isolated and used for establishing the absolute calibration of the instrument on one hand, and for tracking the inevitable changes in this calibration during the mission on the other hand. This observation strategy is expected to mitigate the challenges currently faced by SWAN.

\subsubsection{Interstellar Probe} \label{sec:isp}

Interstellar Probe is a heliophysics mission concept to fly into the nearby interstellar space through the heliosphere boundary with the science goal to “Understand our habitable astrosphere and its home in the galaxy”. Interstellar Probe will investigate (1) the fundamental physical processes which form the heliosphere boundary and uphold the vast heliosphere, (2) the global dynamics of the heliosphere driven by the Sun's activity and possible LISM structures, (3) characteristics of the LISM. Understanding the properties of interstellar H atoms and ion-neutral interaction processes from the Sun to the LISM span all the science objectives of the mission.

An Interstellar Probe mission with a UV spectrograph on board will answer the compelling science questions by making high-spectral-resolution ($\sim 5-10$ km/s) measurements of scattered Lyman-$\alpha$ emission in different lines of sight on outward trajectory from the Sun reaching distances up to 400 AU. The spectral measurements will enable
(1) constraining the hydrogen velocity distribution function throughout the heliosphere and beyond in the LISM,
(2) discovering a position of the hydrogen wall and inferring the 3D structure of this unique unexplored global feature,
(3) determining the properties of hot H atoms created in the heliosheath and their spatial variations (enabling an independent diagnostic of the global heliosheath structure),
(4) determining a deflection of interstellar hydrogen flow in the heliosphere compared to pristine interstellar flow and discovering any deviations from the previously reported deflection of 4$^\circ$ \citep{lallement2005, lallement2010}, and
(5) identifying Galactic and extragalactic components of Lyman-$\alpha$ \citep{lallement2011, katushkina2017, gladstone2018, gladstone2021}. 

Spectra taken at different distances from the Sun and in different LOS will for the first time enable global diagnostics of the non-Maxwellian velocity distribution function of the interstellar hydrogen and therefore understanding of plasma–hydrogen coupling processes in the context of the global heliosphere–LISM interaction. 
To infer spatial variations of LOS hydrogen velocity distributions, multiple look directions are required, in particular toward the nose, toward the tail of the heliosphere, and sidewise, covering at least half of the sky. A measurement cadence on the order of a few months would be sufficient to investigate possible variations of Lyman-$\alpha$ spectra within a year due to solar effects \citep{quemerais2006b}. 

UV instruments that are capable of resolving a line profile of backscattered Lyman-$\alpha$ emission include a high-resolution spectrograph and a spatial heterodyne spectrometer (SHS). MAVEN/IUVS is an example of a spectrograph that includes a far-UV spectral channel that uses an echelle grating to resolve H and D Lyman-$\alpha$ lines (echelle channels were also implemented in the HST/GHRS and HST/STIS). These instrument packages are typically large. A reduced-mass compact spectrograph with a high spectral resolution and sensitivity is a subject of the future technology development for a UV instrument on the Interstellar Probe mission. 

An alternative to grating spectrographs would be an SHS with a high resolving power and a compact design which recently has been under development for laboratory tests and sounding rocket flights \citep{harris2018, hosseini2019}. At the time of this publication, SHS has not yet demonstrated a capability to measure a Lyman-$\alpha$ line profile in space.

\begin{acknowledgements}
This work was made possible by the International Space Science Institute and its interdisciplinary workshop ``The Heliosphere in the Local Interstellar Medium'', URL: \url{www.issibern.ch/workshops/heliosphere}.\\
M.B. was supported by Polish National Science Centre grant 2019/35/B/ST9/01241.
E.P. was supported by NASA Task Order NNN06AA01C.
\end{acknowledgements}


\section*{Conflict of interest}
The authors declare that they have no conflict of interest.

\section*{Acronyms}
AU -- astronomical unit,
ENA -- energetic neutral atom,
GHRS -- Goddard High Resolution Spectrograph,\\
GLOWS -- GLObal solar Wind Structure,
HST -- Hubble Space Telescope,
IBEX -- Interstellar Boundary Explorer,
IMAP -- Interstellar Mapping and Acceleration Probe,
IP -- interplanetary,
IPH -- interplanetary hydrogen,
IPS -- interplanetary scintillation,
IsMF -- interstellar magnetic field,
ISM -- interstellar medium,
ISN – interstellar neutral,
IUE -- International Ultraviolet Explorer,
IUVS -- Imaging Ultraviolet Spectrograph,
LISM -- local interstellar medium,
LOS -- line of sight,
MAVEN -- Mars Atmosphere and Volatile EvolutioN,
MHD -- magnetohydrodynamic,
OGO-5 -- Orbiting Geophysical Observatory number 5,
SHS -- spatial heterodyne spectrometer,
SOHO -- Solar and Heliospheric Observatory,
STIS -- Space Telescope Imaging Spectrograph,
SW -- solar wind,
SWAN -- Solar Wind ANisotropy,
UV -- ultraviolet,
UVS -- Ultraviolet Spectrometer.



\bibliographystyle{spbasic}         

\bibliography{ISSI_Lyman_alpha}

\end{document}